\documentclass[journal=nalefd, manuscript=letter,layout=twocolumn]{achemso}

\usepackage[acronym]{glossaries}
\usepackage{amsmath}
\usepackage[colorlinks, allcolors=blue]{hyperref}

\newacronym{2D}{2D}{two-dimensional}
\newacronym{QW}{QW}{quantum well}

\newacronym{LED}{LED}{light-emitting diode}
\newacronym{RT}{RT}{room temperature}
\newacronym{vdW}{vdW}{van der Waals}
\newacronym{TMD}{TMD}{transition-metal dichalcogenide}
\newacronym{NRC}{NRC}{non-radiative recombination center}
\newacronym{FRET}{FRET}{Förster resonance energy transfer}
\newacronym{CL}{CL}{cathodoluminescence}
\newacronym{1L}{ML}{monolayer}
\newacronym{ZPL}{ZPL}{zero-phonon line}
\newacronym{LO}{LO}{longitudinal optical}
\newacronym{ST}{ST}{surface trap}
\newacronym{uGaN}{$u$-GaN}{unintentionally doped GaN}
\newacronym{DoS}{DoS}{density of states}
\newacronym{PL}{PL}{photoluminescence}
\newacronym{DA}{$d_{\text{DA}}$}{donor-acceptor distance}

\newacronym{MOVPE}{MOVPE}{metalorganic vapor phase epitaxy}
\newacronym{FS}{FS}{free-standing}
\newacronym{LT}{LT}{low-temperature}
\newacronym{AFM}{AFM}{atomic force microscopy}
\newacronym{PC}{PC}{poly(bisphenol A carbonate)}
\newacronym{PDMS}{PDMS}{polydimethylsiloxane}
\newacronym{FWHM}{FWHM}{full width at half maximum}
\newacronym{BL}{BL}{blue luminescence}
\newacronym{RRR}{$R_{\text{r}}$}{radiative recombination rate}
\newacronym{ERR}{$R_{\text{eff}}$}{effective recombination rate}
\newacronym{d_eff}{$d_{\text{eff}}$}{effective thickness}
\newacronym{cw}{cw}{continuous wave}

\usepackage{xcolor}
\usepackage{wrapfig}
\usepackage{chemformula} 
\usepackage[T1]{fontenc} 
\usepackage{booktabs}
\usepackage{caption}
\author{Danxuan Chen}
\affiliation[LASPE]
{Laboratory of Advanced Semiconductors for Photonics and Electronics (LASPE)}
\email{danxuan.chen@epfl.ch}
\phone{+41 (0)21 693 45 33}

\author{Jin Jiang}
\affiliation[LQP]
{Laboratory of Quantum Physics (LQP) \\ \'Ecole Polytechnique F\'ed\'erale de Lausanne (EPFL), CH-1015 Lausanne, Switzerland}

\author{Thomas F. K. Weatherley}
\affiliation[LASPE]
{Laboratory of Advanced Semiconductors for Photonics and Electronics (LASPE)}

\author{Jean-François Carlin}
\affiliation[LASPE]
{Laboratory of Advanced Semiconductors for Photonics and Electronics (LASPE)}

\author{Mitali Banerjee}
\affiliation[LQP]
{Laboratory of Quantum Physics (LQP) \\ \'Ecole Polytechnique F\'ed\'erale de Lausanne (EPFL), CH-1015 Lausanne, Switzerland}

\author{Nicolas Grandjean}
\affiliation[LASPE]
{Laboratory of Advanced Semiconductors for Photonics and Electronics (LASPE)}

\title[An \textsf{achemso} demo]
  {Excitonic interplay between surface polar III-nitride quantum wells and MoS\textsubscript{2} monolayer}


\begin{document}

\pdfbookmark[1]{Abstract}{sec:abs}
\begin{abstract}

	III-nitride wide bandgap semiconductors exhibit large exciton binding energies, preserving strong excitonic effects at room temperature. 
	On the other hand, semiconducting \acrfull{2D} materials, including MoS\textsubscript{2}, also exhibit strong excitonic effects, attributed to enhanced Coulomb interactions.
	This study investigates excitonic interactions between surface GaN \acrfull{QW} and \acrshort{2D} MoS\textsubscript{2} in van der Waals heterostructures by varying the spacing between these two excitonic systems.
	Optical property investigation first demonstrates the effective passivation of defect states at the GaN surface through MoS\textsubscript{2} coating. 
	Furthermore, a strong interplay is observed between MoS\textsubscript{2} monolayers and GaN \acrshort{QW} excitonic transitions. 
	This highlights the interest of the \acrshort{2D} material/III-nitride \acrshort{QW} system to study near-field interactions, such as \acrlong{FRET}, which could open up novel optoelectronic devices based on such hybrid excitonic structures.

\end{abstract}

\section{Introduction}
\pdfbookmark[1]{Introduction}{sec:intro}

	\hspace{\parindent}Since the breakthroughs of blue \glspl{LED} in the 1990s \cite{Nakamura1994}, III-nitrides have emerged as a major family of contemporary semiconductors.
	This development is attributed to their remarkable optoelectronic properties, including a direct bandgap spanning from the near infrared to the deep ultraviolet, thermal and chemical robustness, and large exciton binding energies \cite{Steube1997}, which ensures strong excitonic effects \cite{Malpuech2002}, with for instance, the achievement of \gls{RT} polariton lasing \cite{Christopoulos2007}.

	The isolation of graphene in 2004 \cite{Novoselov2004} marked the inception of a new era in solid-state  physics. 
	Owing to the weak interlayer \gls{vdW} interactions, layered materials can be easily exfoliated into atomically thin layers, \latin{i.e.}, \gls{2D} materials, and seamlessly integrated with other materials, granting notable flexibility in constructing \gls{vdW} heterostructures.
	Among various \gls{2D} materials, semiconducting \glspl{TMD}, such as MoS\textsubscript{2}, exhibit a sizeable bandgap \cite{Chaves2020} and strong light-matter coupling \cite{Britnell2013}, making them highly desirable for optoelectronic applications. 
	When the layer thickness is reduced to the atomic scale, carriers in \gls{2D} \glspl{TMD} exhibit excitons with binding energies typically one to two orders of magnitude larger than those observed in conventional semiconductors \cite{Mueller2018},  which ensures robust excitonic features at \gls{RT} \cite{Ciarrocchi2022}.
	
	Mixed-dimensional \gls{vdW} heterostructures combining \glspl{TMD} with III-nitrides have already been proposed for a diverse range of applications including \glspl{LED} \cite{Li2015}, water splitting \cite{Zhang2018}, and photodetection \cite{Jain2020}. 
	In such heterostructures, III-nitrides are typically utilized in bulk form.
	A closer look at \gls{TMD}/III-nitride interactions requires the study of the exitonic interplay at the \gls{vdW} interface only. 
	However, conventional semiconductor surfaces possess deep states in the bandgap which act as \glspl{NRC}. 
	These \glspl{NRC} can effectively suppress excitonic features near the surface \cite{Chang1993}.
	In contrast to other III-V semiconductors, III-nitrides exhibit relatively low surface recombination velocity \cite{Karpov2016}. 
	In addition, by carefully designing the \gls{QW} structure, the exciton binding energies can be further increased compared to those in bulk materials \cite{Grandjean1999}. 
	Moreover, wurtzite III-nitrides exhibit large polarization mismatches at heterointerfaces \cite{Bernardini1997}, which results in a strong built-in electric field in the well and thereby large electron-hole dipoles in the \gls{QW}. 
	Then, \gls{2D} \gls{TMD} excitons in their vicinity should enable dipole-dipole coupling, known as \gls{FRET} \cite{Forster1960}. 
	This effect has been observed between CdSe/ZnS nanocrystals and surface InGaN \glspl{QW} and could be utilized for developing highly efficient “energy-transfer color converters” \cite{Achermann2004}.

	In this study, we tune the excitonic interaction between \gls{2D} MoS\textsubscript{2} and surface polar GaN/AlGaN \glspl{QW} by varying the distance, $d$, between the two.
	\Gls{CL} measurements on bare \glspl{QW} show appreciable emission at \gls{RT} even in the absence of a surface barrier, which confirms a low surface recombination rate.  
	Upon coating the surface with MoS\textsubscript{2}, the \gls{CL} intensity markedly changes, indicating that deep traps still exist at the surface and are passivated by the MoS\textsubscript{2} coating. 
	Eventually, \gls{CL} results from the sample with $d=15$~nm unveil a pronounced change of the emission at \gls{QW} excitonic transitions upon \gls{1L}-MoS\textsubscript{2} deposition. 
	This phenomenon points out the role of \gls{FRET} between \gls{1L}-MoS\textsubscript{2} and surface GaN \gls{QW}.

\section{Results and Discussion}
\pdfbookmark[1]{Results and Discussion}{sec:results}

\begin{figure*}[t]
	\centering
	\includegraphics[width=1\textwidth]{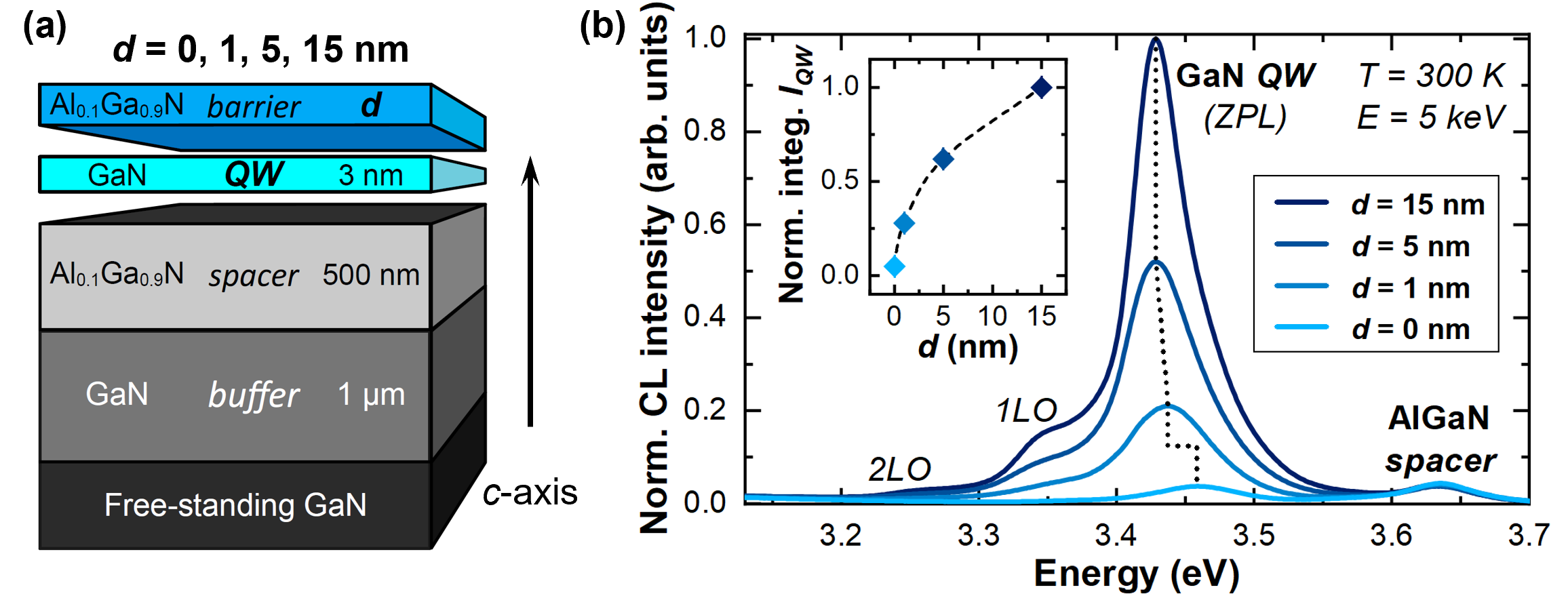}
	\caption{
		\textbf{(a)} Sample structure of surface \glspl{QW} and
		\textbf{(b)} corresponding \gls{CL} spectra acquired at 300~K under an electron beam energy of 5~keV.
		In the inset, the integrated \gls{QW} intensity ($I_{\text{QW}}$), including the \gls{ZPL} and its \gls{LO} phonon replicas, is plotted as a function of surface barrier thickness ($d$).
		The intensity error bars are not visible in the plot as they are smaller than the size of the diamond symbol used.
	}
	\label{QWs}
\end{figure*}

	\hspace{\parindent}The sample structure of $c$-axis oriented surface GaN/AlGaN \glspl{QW} is illustrated in Fig.~\ref{QWs}\textbf{(a)}. 
	The well thickness of 3~nm is chosen as a compromise between maximizing the binding energy and maintaining the dipolar nature of excitons \cite{Grandjean1999}. 
	The sample without the AlGaN surface barrier is considered a “\gls{QW}” with $d=0$~nm. 
	A 500~nm thick AlGaN spacer is inserted beneath the \gls{QW} to prevent electron-beam-generated carriers from reaching the buffer where they could give rise to parasitic luminescence (Supporting Information (SI) Sec.~\hyperref[CASINO_sec]{2}).
	The \gls{CL} spectra of surface \glspl{QW} are presented in Fig.~\ref{QWs}\textbf{(b)}, with dominant peaks at $\sim3.43$~eV and $\sim3.63$~eV, attributed to the GaN \glspl{QW} and AlGaN spacers, respectively.
	The first important observation is that, in contrast to near-surface GaAs \glspl{QW} \cite{Chang1993}, all present GaN \glspl{QW} exhibit a strong emission at \gls{RT}, even in the absence of a surface barrier (SI Sec.~\hyperref[QWs_sec]{3}).

	Now we will delve into the characteristics of each \gls{CL} peak. 
	As the thickness of the surface barriers is smaller than the carrier diffusion length in Al\textsubscript{0.1}Ga\textsubscript{0.9}N \cite{Gonzalez2001}, it is reasonable to assume that most of the carriers generated in the surface barrier diffuse to the \gls{QW} where they recombine.
	Hence, the AlGaN emission in Fig.~\ref{QWs}\textbf{(b)} originates from the spacer that is farther away from the surface, where carriers recombine before reaching the \gls{QW}.
	The depth of the interaction volume of the 5~keV electron beam in these samples is more than 100~nm (SI Sec.~\hyperref[CASINO_sec]{2}), which implies that the total carrier generation rate remains nearly constant regardless of the position of the \gls{QW} in the surface region, \latin{i.e.}, $d$.
	This is corroborated by the comparable AlGaN peak intensity in all samples (SI Sec.~\hyperref[QWs_sec]{3}).
	On the other hand, the GaN \gls{QW} peak changes markedly with $d$. 
	In contrast to other surface \glspl{QW}, the peak of the uncapped well ($d=0$~nm) blueshifts by $\sim30$~meV. 
	This can be explained by a stronger carrier quantum confinement imposed by the free surface (SI Sec.~\hyperref[QWs_sec]{3}).
	The integrated \gls{QW} intensity (see SI Sec.~\hyperref[QWs_sec]{3} for calculation details) in Fig.~\ref{QWs}\textbf{(b)} exhibits a nonlinear increase with increasing $d$, which indicates a reduction of non-radiative recombination channels.
	For a $c$-plane (Al)GaN surface, a high density of deep levels can act as effective \glspl{NRC} \cite{VdW2007}. 
	Thus, carriers in surface \glspl{QW} can tunnel through the nanoscale barrier and be captured by \glspl{ST}. 
	The $d$-dependent \gls{QW} intensity can be modeled by an exponential function to account for carrier tunneling \cite{Chang1993} (SI Sec.~\hyperref[QWs_sec]{3}). 
	Therefore, the nonlinear increase in \gls{QW} emission with increasing $d$ demonstrates the significant impact of \glspl{ST} on surface GaN \glspl{QW}, despite the typically low surface recombination velocity usually ascribed to III-nitrides. 
	This highlights the importance of surface passivation for III-nitrides, particularly in devices with a high surface-to-volume ratio \cite{Seong2020}.


\begin{figure*}[t]
	\centering
	\includegraphics[width=1\textwidth]{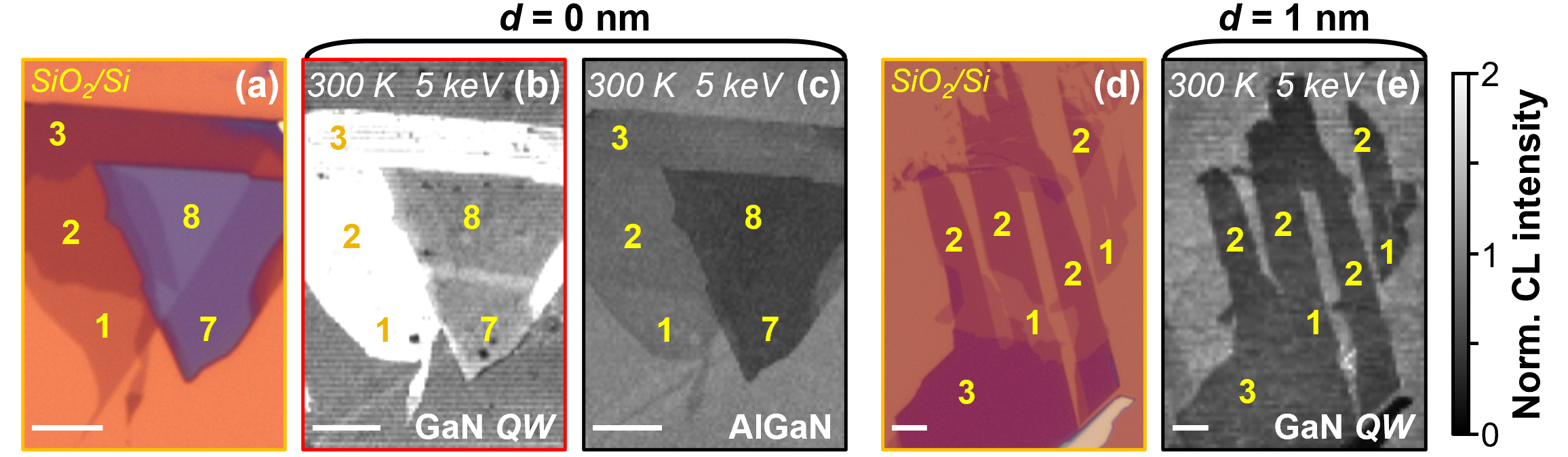}
	\caption{
		\textbf{(a,~d)} Optical micrographs of the selected MoS\textsubscript{2} flakes on a SiO\textsubscript{2}/Si substrate.
		Normalized integrated \gls{CL} intensity maps of the \textbf{(b)} GaN \gls{QW} and \textbf{(c)} AlGaN emissions from the uncapped GaN \gls{QW} ($d=0$~nm), as well as \textbf{(e)} the GaN \gls{QW} emission from the sample with $d=1$~nm, acquired with an electron beam energy of 5~keV at 300~K. 		
		For each map, the normalization was performed using the average intensity in the region without MoS\textsubscript{2}.
		All \gls{CL} maps are plotted on an intensity scale of $0-2$.
		The numbers in yellow indicate the number of MoS\textsubscript{2} \glspl{1L} in the corresponding region.
		Scale bars correspond to a length of 5~$\mu$m.
	}
	\label{CLmaps}
\end{figure*}

	Mechanically exfoliated MoS\textsubscript{2} flakes were prepared on a SiO\textsubscript{2}/Si substrate, where the contrast in an optical microscope is highly sensitive to MoS\textsubscript{2} thickness due to light interference \cite{Li2013} (Figs.~\ref{CLmaps}\textbf{(a,~d)}).
	After precise characterization of the layer thickness (SI Sec.~\hyperref[MoS2_thickness_sec]{4}), the selected flakes were deposited on the surface GaN \glspl{QW}.
	Hyperspectral \gls{CL} maps were acquired on the MoS\textsubscript{2} flake regions.
	Each pixel in the map corresponds to the \gls{CL} at that position and the associated spectrum was fitted to generate integrated intensity maps of the GaN \gls{QW} and AlGaN emissions (SI Sec.~\hyperref[CL_processing_sec]{5}). 
	To facilitate the comparison, all intensity maps were normalized by the average intensity of the background (SI Sec.~\hyperref[CL_processing_sec]{5}).	
	
	Among all intensity maps, the uncapped GaN \gls{QW} ($d=0$~nm) shows a peculiar behavior: the region covered by MoS\textsubscript{2} exhibits a strongly enhanced emission (Fig.~\ref{CLmaps}\textbf{(b)}). 
	This is not consistent with the high spectral absorptance of MoS\textsubscript{2} in the range of (Al)GaN emission ($\sim10\%$ of the incident light is absorbed by \gls{1L}-MoS\textsubscript{2}) \cite{Dumcenco2015}. 
	In contrast, the AlGaN intensity map extracted from the same sample shows a decrease in intensity with increasing MoS\textsubscript{2} thickness (Fig.~\ref{CLmaps}\textbf{(c)}), as expected from absorption \cite{Castellanos-Gomez2016}. 
	Additionally, when the GaN \gls{QW} is slightly moved away from the surface, \latin{i.e.}, $d=1$~nm, its \gls{CL} intensity is also reduced with the presence of MoS\textsubscript{2} (Fig.~\ref{CLmaps}\textbf{(e)}).
	It is noteworthy that, despite similar MoS\textsubscript{2} spectral absorptance in the corresponding energy ranges, the contrast observed from the GaN \gls{QW} with $d=1$~nm differs from that of the AlGaN emission (Fig.~\ref{CLmaps}\textbf{(c)}).
	The former shows a more pronounced reduction in intensity throughout the entire region covered by MoS\textsubscript{2} of $1-3$~\glspl{1L}. 
	The underlying reason will be elucidated later.
	On the other hand, considering the abnormal increase in \gls{CL} intensity linked to MoS\textsubscript{2} for the uncapped \gls{QW}, it is evident that only the emission of the layer in direct contact with the MoS\textsubscript{2} flake is enhanced, indicating that the MoS\textsubscript{2}-enhanced GaN emission is associated with surface passivation.

	To confirm this hypothesis, we performed \gls{CL} experiments on a GaN epilayer coated by MoS\textsubscript{2} (SI Sec.~\hyperref[MoS2_on_buffer_sec]{6}). 
	The \gls{CL} map also exhibits an increase in GaN emission in the presence of MoS\textsubscript{2}, albeit weaker compared to the case of the uncapped GaN \gls{QW}. 
	This is consistent with a surface passivation effect: in a GaN epilayer, \gls{CL} emission comes from both the surface and bulk regions, which diminishes the surface impact.
	
	\begin{figure*}[t]
		\centering
		\includegraphics[width=1\textwidth]{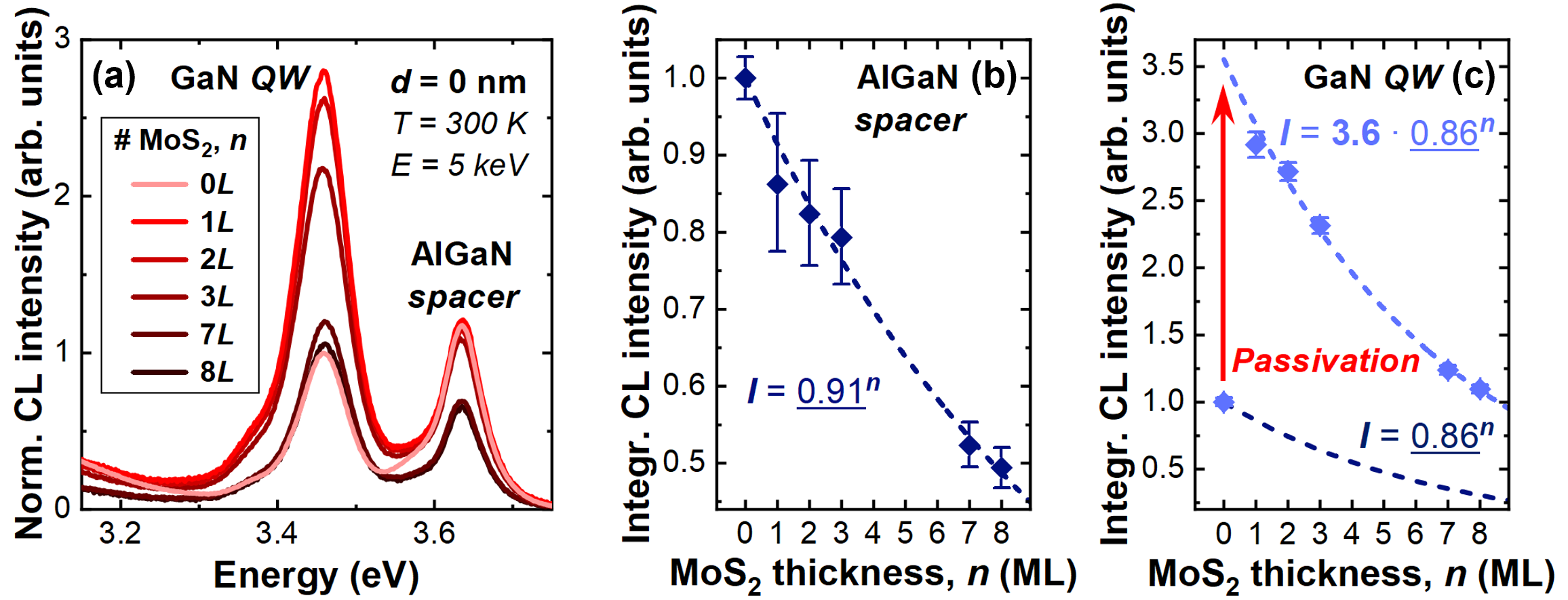}
		\caption{
			\textbf{(a)} Average \gls{RT} \gls{CL} spectra collected from the uncapped GaN \gls{QW} ($d=0$~nm) in areas with varying MoS\textsubscript{2} thickness, represented by the number of \glspl{1L} ($n$). 
			All spectra are normalized to the peak intensity of the GaN \gls{QW} emission in the region without MoS\textsubscript{2} ($0L$).
			Normalized integrated \gls{CL} intensity as a function of MoS\textsubscript{2} thickness for \textbf{(b)} the AlGaN spacer emission and \textbf{(c)} the GaN \gls{QW} emission. 
			The dashed lines represent the fit assuming the same absorption in each \gls{1L}-MoS\textsubscript{2}, with the corresponding expression next to them.
			Some intensity error bars are not visible in the plots as they are smaller than the size of the diamond symbol used.
		}
		\label{0nm}
	\end{figure*}
	
	In order to obtain more quantitative information from the results of the uncapped \gls{QW}, average \gls{CL} spectra were extracted from regions with different MoS\textsubscript{2} thicknesses (SI Sec.~\hyperref[CL_processing_sec]{5}), as shown in Fig.~\ref{0nm}\textbf{(a)}. 
	These spectra were deconvoluted to derive the integrated intensities of the GaN \gls{QW} and AlGaN spacer peaks, respectively (SI Sec.~\hyperref[QWs_sec]{3}).
	Let us consider first the AlGaN spacer emission, which is not affected by any surface effects (Fig.~\ref{0nm}\textbf{(b)}).
	It exhibits a monotonic decrease with increasing MoS\textsubscript{2} thickness, as expected from absorption.
	It is important to note that the presence of MoS\textsubscript{2} has a negligible impact on carrier injection into the samples, primarily due to the limited interaction of the electron beam with these ultra-thin layers \cite{Negri2020}.
	To model the AlGaN intensity decrease, we consider that the interlayer coupling in MoS\textsubscript{2} does not strongly influence the absorption, thus the absorption in each \gls{1L} is nearly the same. 
	Therefore, the normalized intensity can be fitted with a power function: $I(n) = (1-a)^n$, where $n$ is the number of MoS\textsubscript{2} \glspl{1L} and $a$ is the absorptance in each \gls{1L}. 
	The fit gives $a \approx 10\%$, which agrees with the absorptance measured in \gls{1L}-MoS\textsubscript{2} at 3.63 eV \cite{Dumcenco2015}, \latin{i.e.}, the peak energy of AlGaN emission.
	Similarly, Fig.~\ref{0nm}\textbf{(c)} shows the plot of the GaN \gls{QW} \gls{CL} intensity as a function of MoS\textsubscript{2} thickness.
	Fitting the data with the same absorption model reproduces the overall trend, with $a \approx 10\%$, matching \gls{1L}-MoS\textsubscript{2} absorptance at 3.45 eV \cite{Dumcenco2015}.
	However, this fit does not capture the experimental data at $n=0$, instead predicting an intensity 3.6 times higher than the measured value.
	This indicates that the deposition of the first \gls{1L}-MoS\textsubscript{2} results in a strong increase in the surface emission, \latin{i.e.}, the emission of the uncapped \gls{QW}, due to single \gls{1L}-MoS\textsubscript{2} surface passivation effect. 
	
	\glsreset{ST}
	\begin{figure*}[t]
		\centering
		\includegraphics[width=1\textwidth]{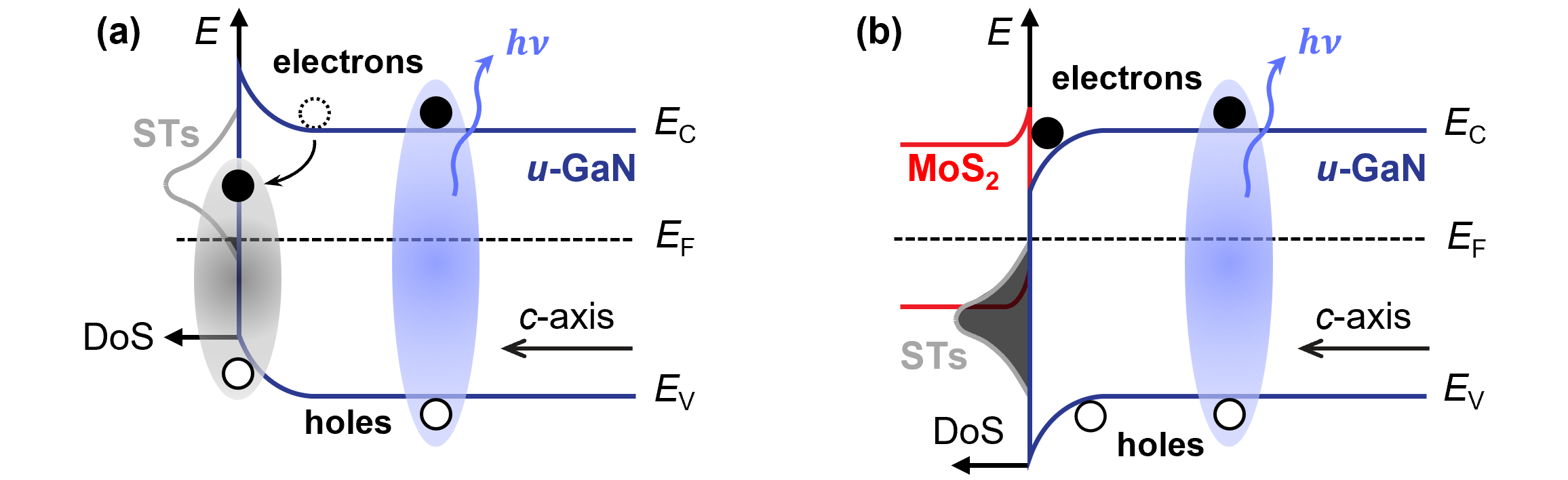}
		\caption{
			Surface band bending of \textbf{(a)} bare \gls{uGaN} and \textbf{(b)} \gls{uGaN} coated with MoS\textsubscript{2} after interfacial charge transfer \cite{Jain2020}.
			The \gls{DoS} of \glspl{ST} in \gls{uGaN} is depicted in both situations, with the black shaded part representing occupied states.
			The grey ellipse represents an exciton captured by a \gls{ST} and does not emit light, while the blue ellipses represent free excitons that can emit light through radiative recombination.
		}
		\label{Passivation}
	\end{figure*}
	
	A possible explanation for surface passivation is based on type-II band alignment between MoS\textsubscript{2} and GaN \cite{Jain2020}: 
	the \acrlong{uGaN} surface exhibits an upward surface band bending due to the presence of \glspl{ST} (Fig.~\ref{Passivation}\textbf{(a)}); 
	with MoS\textsubscript{2} deposited on the surface, charge transfer between the two materials leads to an upward band bending in MoS\textsubscript{2} and a downward band bending in GaN (Fig.~\ref{Passivation}\textbf{(b)}). 
	Consequently, \glspl{ST} are occupied and are no longer capable of trapping excitons in the surface region, therefore, surface emission is markedly enhanced.
	
\begin{figure*}[t]
		\centering
		\includegraphics[width=1\textwidth]{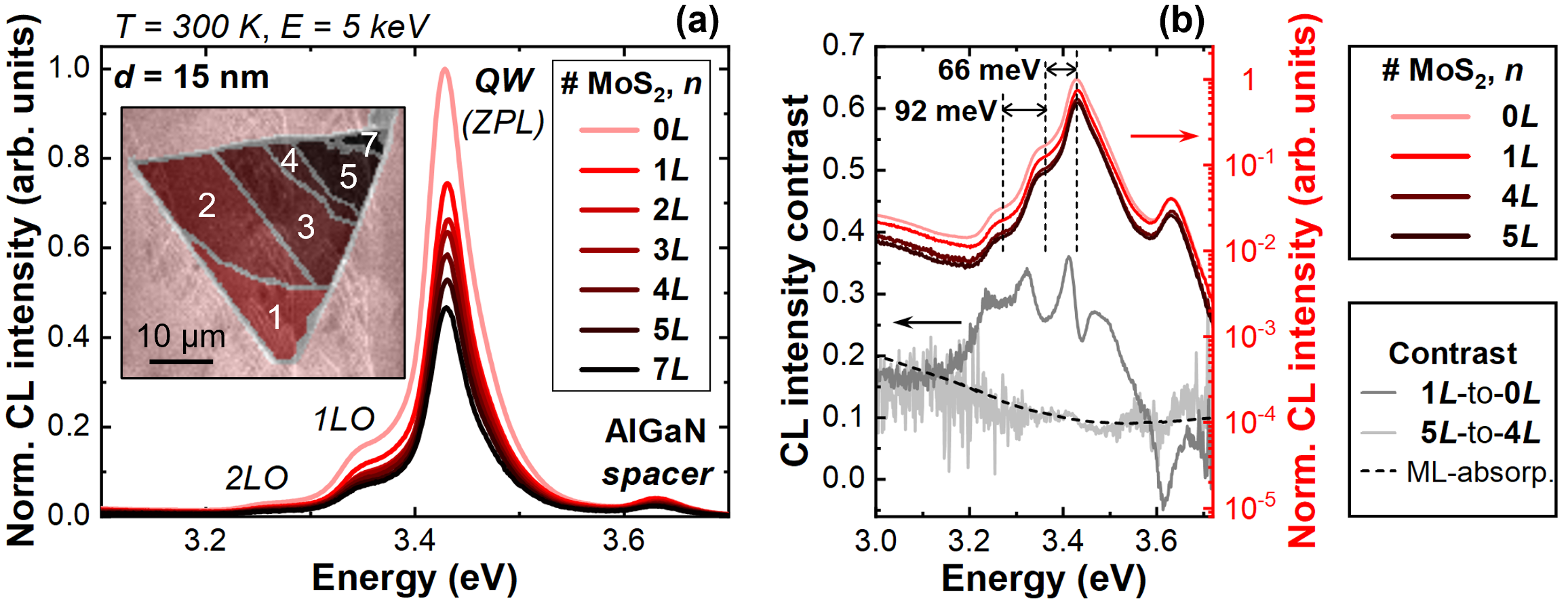}
		\caption{
			\textbf{(a)} Average \gls{RT} \gls{CL} spectra taken on the sample with $d=15$~nm in regions with different MoS\textsubscript{2} thicknesses. 
			The corresponding regions are shown in the inset. 
			\textbf{(b)} Intensity contrast between the background ($0L$) and $1L$-MoS\textsubscript{2} region, and contrast between the $4L$- and $5L$-MoS\textsubscript{2} regions, in comparison to the reported \gls{1L}-MoS\textsubscript{2} spectral absorptance \cite{Dumcenco2015}. 
			The related \gls{CL} spectra are plotted to illustrate the energy range of different emissions.}
		\label{15nm}
	\end{figure*}

	As the deposition of MoS\textsubscript{2} exerts a significant impact on the emission of the uncapped \gls{QW}, it is not a suitable sample for studying excitonic interplay between the two materials. 
	Conversely, in the sample with $d=15$~nm, carriers confined in the well are no longer interacting with \glspl{ST} (SI Sec.~\hyperref[QWs_sec]{3}), making this system ideal for gaining insight into the excitonic interactions.~The \gls{CL} spectra with different MoS\textsubscript{2} thicknesses are displayed in Fig.~\ref{15nm}\textbf{(a)}.
	Throughout the entire detection range, from $\sim3.0$~eV to $\sim3.7$~eV, the \gls{CL} intensity decreases with increasing MoS\textsubscript{2} thickness. 
	This energy range coincides with the C-absorption band of \gls{1L}-MoS\textsubscript{2} \cite{Dumcenco2015}.
	Interestingly, for all peaks related to the GaN \gls{QW} emission, including the \gls{ZPL} and its \gls{LO} phonon replicas, the intensity difference between \gls{1L}-MoS\textsubscript{2} capped QW ($1L$) and bare QW ($0L$) is notably larger than the intensity change observed when increasing the MoS\textsubscript{2} thickness by 1~\gls{1L}.
	
	To get a deeper insight into the influence of MoS\textsubscript{2} on the \gls{QW} emission, we computed the intensity contrast, following the formula: 
	\begin{equation*}
		\text{Contrast}[(n+1)L\text{-to-}nL](E) = 1-\frac{I_{(n+1)L} (E)}{ I_{nL} (E) }
	\end{equation*}
	where $I_{nL}$ is the \gls{CL} intensity of the region covered by $nL$-MoS\textsubscript{2}, \latin{i.e.}, MoS\textsubscript{2} of $n$~\glspl{1L}, as a function of photon energy, $E$.
	As depicted in Fig.~\ref{15nm}\textbf{(b)}, the intensity contrast between regions covered by $4L$- and $5L$-MoS\textsubscript{2} (light grey curve) aligns with the reported \gls{1L}-MoS\textsubscript{2} spectral absorptance \cite{Dumcenco2015} (black dashed curve). 
	On the contrary, the contrast between the background and the $1L$-MoS\textsubscript{2} region (dark grey curve) exhibits a significant discrepancy in the emission range of GaN \gls{QW} and AlGaN. 
	In the range around the AlGaN bandgap, a peak with a negative contrast is observed, \latin{i.e.}, higher AlGaN emission in the presence of \gls{1L}-MoS\textsubscript{2}.
	This can be explained by the surface passivation induced by \gls{1L}-MoS\textsubscript{2}, which enhances surface emission from the AlGaN barrier (SI Sec.~\hyperref[15nm_sec]{7}).
	In the range of GaN \gls{QW} emission, multiple peaks, with contrasts much higher than expected from absorptance only, are observed. 
	The dominant peak at 3.41~eV, along with the two lower-energy peaks, can be associated with the \gls{ZPL} and the corresponding \gls{LO} phonon replicas of the GaN \gls{QW} emission in the \gls{CL} spectrum (light red curve).
	The carrier density in the well is on the order of $10^{12}\text{~cm}^{-2}$ (SI Sec.~\hyperref[CASINO_sec]{2}). 
	This is close to the critical carrier density of the Mott transition reported for similar GaN \glspl{QW} \cite{Rossbach2014, Shahmohammadi2014}. 
	Therefore, \gls{CL} peaks of the GaN \gls{QW} can arise from two possible physical origins: excitonic transitions or electron-hole plasma emission.
	The presence of excitonic features in the well can be confirmed by the energy spacing between adjacent peaks of the \gls{QW} emission.
	Theoretically, the peak emission energy of the $m$\textsuperscript{th} \gls{LO} phonon replica of excitons can be expressed as \cite{Klingshirn2005}:
	\begin{equation*}
		\begin{split}
			E_m &= E_0 - m E_{\text{LO}} + E_{\text{kin}} \\
			&= E_0 - m E_{\text{LO}} + \frac{D}{2} k_B T
		\end{split}
	\end{equation*}
	where $E_0$ is the peak energy of the \gls{ZPL}, $E_{\text{LO}}$ is the \gls{LO} phonon energy (92~meV for GaN \cite{Hofstetter2020}), and $E_{\text{kin}}$ is the kinetic energy of the excitons, which depends on $D$, the dimensionality of the system ($D=2$ for excitons confined in a \gls{QW}).
	$T$ is the lattice temperature, and $k_B$ is the Boltzmann constant. 
	For excitonic emission from a GaN/AlGaN \gls{QW} at 300~K, the theoretical energy spacing between the \gls{ZPL} and the 1\textsuperscript{st} \gls{LO} phonon replica should be $E_{\text{LO}} - k_B T$, \latin{i.e.}, 66~meV, whereas the energy spacing between the 1\textsuperscript{st} and 2\textsuperscript{nd} \gls{LO} phonon replicas should be  $E_{\text{LO}}$, \latin{i.e.}, 92~meV.
	This agrees perfectly with the \gls{CL} spectra shown in Fig.~\ref{15nm}\textbf{(b)}, which confirms the excitonic nature of the \gls{QW} emission.
	Furthermore, the \gls{CL} emission profile of the GaN \glspl{QW} is compared to the emission measured by \gls{PL} using a \gls{cw} 325~nm laser with a power density of $\sim16$~W/cm\textsuperscript{2}, which results in much lower carrier injection into the \glspl{QW} (SI Sec.~\hyperref[QWs_sec]{3}). 
	Although the energy peak blueshifts in \gls{CL}, which may be attributed to the screening of the built-in field at higher carrier density, the line shape remains nearly unchanged. 
	This observation excludes Fermi filling of continuum states, characterized by a strong extension of the high-energy tail, which is typically observed beyond the Mott transition \cite{Rossbach2014}.
	Therefore, our findings demonstrate that, besides the conventional absorption, there is an additional energy transfer channel between the GaN \gls{QW} and \gls{1L}-MoS\textsubscript{2}, in resonance with the excitonic transitions in the \gls{QW}. 
	Such an effect is clearly observed in the contrast induced by the direct deposition of \gls{1L}-MoS\textsubscript{2} on the \gls{QW} surface. 

	The relevant mechanism is probably \gls{FRET}, which provides an efficient energy transfer between energy “donor” and “acceptor” materials through non-radiative dipole–dipole coupling \cite{Forster1960}. 
	In our case, carriers confined in the polar GaN/AlGaN \gls{QW} are subjected to a built-in field, forming dipoles which act as energy “donors”. 
	On the other hand, different from excitons in other MoS\textsubscript{2} layers, excitons in the \gls{1L}-MoS\textsubscript{2} on the surface are the closest to the \gls{QW} dipoles and experience a weaker dielectric screening due to the absence of a layer on top.
	Furthermore, since the AlGaN surface barrier possesses a polar nature as all III-nitride layers grown along the $c$-axis, it can enhance the dipolar characteristics of the excitons in direct contact with it.
	All these result in strong dipoles in \gls{1L}-MoS\textsubscript{2}, which act as energy “acceptors”. 
	The associated \gls{DA} is $\sim15$~nm, which is on the order of the typical range for \gls{FRET} \cite{Itskos2007, Prins2014, Taghipour2018}.
	Therefore, we attribute the observed sharp peaks with intensity contrasts as high as $\sim36\%$ to the strong dipolar coupling between excitons in the \gls{QW} and in \gls{1L}-MoS\textsubscript{2} in the near-field regime. 
	This effect also explains the difference in \gls{CL} contrast between the AlGaN emission from the uncapped \gls{QW} (Fig.~\ref{CLmaps}\textbf{(c)}) and the GaN \gls{QW} emission from the sample with $d=1$~nm (Fig.~\ref{CLmaps}\textbf{(e)}). 
	Specifically, the AlGaN emission originates from the bulk region where the excitonic feature is weaker and \gls{DA} is larger. 
	In contrast, the GaN \gls{QW} emission with $d=1$~nm is associated with strong excitonic effect in the well, localized very close to MoS\textsubscript{2}. 
	As a result, a significant quenching of the emission is observed due to efficient \gls{FRET}. 
	The peak \gls{QW} intensity contrast induced by \gls{1L}-MoS\textsubscript{2} on the sample with $d=1$~nm is $\sim52\%$ (SI Sec.~\hyperref[1nm_5nm_sec]{8}), much higher than the 36$\%$ observed for $d=15$~nm. 
	This difference is consistent with the strong dependence of \gls{FRET} on \gls{DA}. 
	It is important to note that with a thin surface barrier of 1~nm, the \gls{QW} emission is still strongly influenced by \glspl{ST}  (SI Sec.~\hyperref[QWs_sec]{3}). 
	The deposition of \gls{1L}-MoS\textsubscript{2} should simultaneously cause a negative contrast due to surface passivation. 
	Therefore, the 52$\%$ contrast deduced from \gls{CL} measurements is likely underestimated.

\newpage

\section{Conclusion}
\pdfbookmark[1]{Conclusion}{sec:conc}

	\hspace{\parindent}In summary, we investigated the optical properties of a series of surface GaN/AlGaN \glspl{QW} with variable nanometer-scale surface barrier thickness, $d=0$~to~15~nm. 
	Thanks to the low surface recombination rate, high \gls{CL} intensity was observed, even from the uncapped \gls{QW} ($d=0$~nm). 
	The \gls{QW} intensity increased nonlinearly with increasing $d$, showing the non-negligible impact of deep traps existing at III-nitride surfaces. 
	Using these surface GaN \glspl{QW} as substrates, we deposited MoS\textsubscript{2} flakes of a few \glspl{1L}.
	The presence of MoS\textsubscript{2} enhanced the emission from the uncapped \gls{QW}, demonstrating III-nitride surface passivation using \gls{2D} material coating.
	For the \gls{QW} with $d=15$~nm, unaffected by \glspl{ST}, a strong excitonic interaction between the GaN \gls{QW} and \gls{1L}-MoS\textsubscript{2} is observed. 
	This effect is attributed to a strong dipole-dipole coupling, \latin{i.e.}, \gls{FRET}, between the excitons of the two materials.
	Our results highlight the potential of surface III-nitride \glspl{QW} as a platform for investigating the near-field interplay between \gls{2D} materials and III-nitrides, which could be applied to develop novel optoelectronics based on such hybrid heterostructures.

\newpage


\section{Supporting Information}

The Supporting Information accompanies this paper. Further details on:
\begin{enumerate}
	\item \hyperref[methods_sec]{Experimental methods}
	\item \hyperref[CASINO_sec]{Carrier injection in \gls{CL}}
	\item \hyperref[QWs_sec]{Optical properties of surface GaN \glspl{QW}}
	\item \hyperref[MoS2_thickness_sec]{MoS\textsubscript{2} thickness determination}
	\item \hyperref[CL_processing_sec]{\Gls{CL} data processing}
	\item \hyperref[MoS2_on_buffer_sec]{MoS\textsubscript{2} on bulk GaN epilayer}
	\item \hyperref[15nm_sec]{MoS\textsubscript{2} on GaN \gls{QW}, $d=15$~nm}
	\item \hyperref[1nm_5nm_sec]{MoS\textsubscript{2} on GaN \glspl{QW}, $d=1,5$~nm}
\end{enumerate}


\begin{acknowledgement}
	
	The authors thank Dr. R. Butté (EPFL) for useful discussions.
	The Interdisciplinary Centre for Electron Microscopy (CIME) at EPFL is acknowledged for access to its facilities.
	M. B. acknowledges the support of SNSF Eccellenza grant No.~PCEGP2\_194528, and support from the QuantERA~II Programme that has received funding from the European Union’s Horizon 2020 research and innovation program under Grant Agreement No.~101017733.
	
\end{acknowledgement}


\clearpage
\onecolumn
\doublespacing

\begin{center}
	\textbf{\Large SUPPORTING INFORMATION}
\end{center}

\setcounter{equation}{0}
\setcounter{figure}{0}
\setcounter{table}{0}
\renewcommand{\thefigure}{S\arabic{figure}}
\pdfbookmark[0]{Supplementary information}{sec:suppl}

\section{1. Experimental methods}
\label{methods_sec}

\hspace{\parindent}\textbf{Sample growth.} 
The samples used in this study were grown by \acrlong{MOVPE} in a horizontal Aixtron 200/4 RF-S reactor on commercial $c$-plane \acrlong{FS} GaN substrates with very low dislocation density, typically a few $10^6\text{~cm}^{-2}$.
The growth process can be divided into two parts.
First, a rapid growth of the GaN buffer and a $\sim500$~nm Al\textsubscript{0.1}Ga\textsubscript{0.9}N spacer is conducted at a growth rate of $\sim2~\mu$m/h and a temperature of 1000~°C.
Trimethylgallium and trimethylaluminum are used as precursors, and H\textsubscript{2} is used as carrier gas.
Then, the growth rate is lowered to 60~nm/h at a temperature of 800~°C for the single \gls{QW} region, which includes a 5~nm Al\textsubscript{0.1}Ga\textsubscript{0.9}N barrier, the 3~nm GaN \gls{QW} layer, and the Al\textsubscript{0.1}Ga\textsubscript{0.9}N surface barrier. 
This \gls{LT} growth is intended to mitigate large-scale Al content fluctuations within the barriers, thereby minimizing the inhomogeneous broadening of the \gls{QW} emission \cite{Feltin2007}.
For the \gls{LT} growth, the metalorganic source for gallium is changed to triethylgallium, and the carrier gas is switched to N\textsubscript{2}. 
The entire structure is grown without any intentional doping.

\textbf{Fabrication of MoS\textsubscript{2}-on-(Al)GaN heterostructures.} 
The MoS\textsubscript{2} flakes were obtained through the well-known “scotch-tape” mechanical exfoliation method \cite{Novoselov2004} and deposited on a 10-minute oxygen-plasma-etched SiO\textsubscript{2}/Si substrate.
The precise thickness of the selected MoS\textsubscript{2} flakes was determined by optical microscopy, \gls{AFM} and Raman spectroscopy (SI Sec.~\hyperref[MoS2_thickness_sec]{4}). 
Following the characterization, the selected flakes were picked up and transferred onto the cleaned surface of the (Al)GaN samples (\glspl{QW} and a bulk GaN epilayer) using a dry transfer technique \cite{Wang2013, Castellanos-Gomez2022}. 
Initially, a high-quality uniform stack of \acrlong{PC}/\acrlong{PDMS} was prepared on a glass slide. 
This stack was then mounted on a homemade transfer stage to pick up MoS\textsubscript{2} at 70~°C.
Subsequently, the MoS\textsubscript{2} flake was transferred to the surface of the (Al)GaN sample at 150~°C. 
Finally, the entire sample was immersed in chloroform to clean its surface.

\textbf{Cathodoluminescence spectroscopy.}
\Gls{CL} imaging was conducted using a specialized scanning electron microscope system (Attolight Rosa 4634) with an acceleration voltage of 5~kV. 
A Cassegrain reflective objective was employed to collect the emitted light, which was subsequently directed to a spectrometer equipped with a 600~lines per mm grating with a blaze wavelength of 300~nm. 
The dispersed light was then captured by a cooled charge-coupled device camera, enabling the recording of a full intensity-energy spectrum at each pixel, \latin{i.e.}, hyperspectral imaging.

\section{2. Estimation of the carrier density in the \glspl{QW}}
\label{CASINO_sec}

\begin{figure}[t!]
	\centering
	\includegraphics[width=1\textwidth]{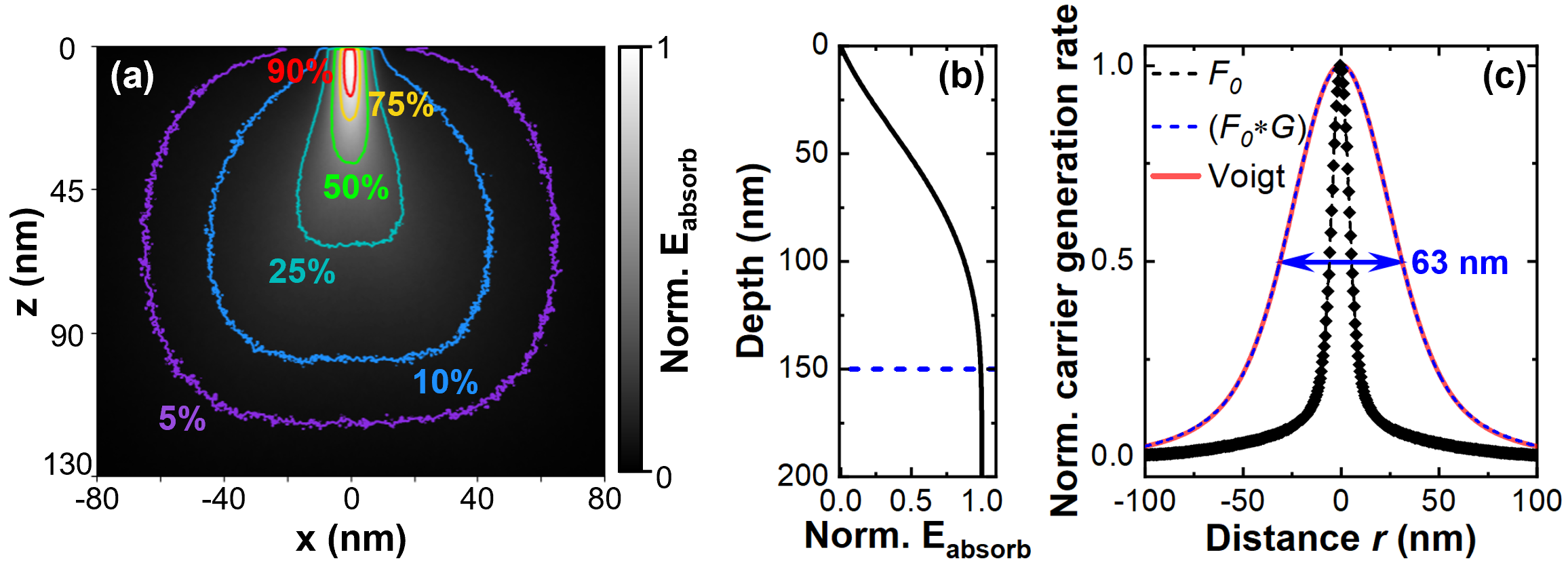}
	\caption{
		\textbf{(a)} Cross-section view of absorbed energy resulting from the Monte Carlo simulation of a 5~keV electron beam interacting with bulk Al\textsubscript{0.1}Ga\textsubscript{0.9}N at 300~K.
		\textbf{(b)} Depth-dependent energy deposition in the sample, normalized by the total absorbed energy. 
		\textbf{(c)} Lateral carrier generation rate distribution, $F(r)$, derived by convolving the simulated carrier generation rate distribution, $F_0(r)$, with a Gaussian function, $G(r)$, characterized by a standard deviation $\sigma=22$~nm, accounting for the broadening due to carrier thermalization at 300~K \cite{Jahn2022}.
		The profile is fitted by a Voigt function with a \acrshort{FWHM} of $\sim63$~nm.}
	\label{Casino}
\end{figure}

\hspace{\parindent}The interaction volume of a 5~keV electron beam at 300~K in bulk Al\textsubscript{0.1}Ga\textsubscript{0.9}N is determined through Monte Carlo simulation (\textit{CASINO}) \cite{Drouin2007}, as depicted in Fig.~\ref{Casino}\textbf{(a)}.
For this simulation, an electron beam containing $1\times10^6$ electrons was used, with a spot size of 25~nm, and an accelerating voltage of 5~kV.
This can be deemed representative of all the samples examined in this study, as the thickness of the GaN \gls{QW} is negligible compared to the overall interaction volume, and a $10\%$ variation in Al content leads to less than $5\%$ change in mass density.
Based on the simulation, the normalized energy deposited in the sample is plotted as a function of depth from the surface (Fig.~\ref{Casino}\textbf{(b)}), which indicates that most of the beam energy is absorbed within a 150~nm region from the surface.
Considering that the minority carrier diffusion length in GaN and Al\textsubscript{0.1}Ga\textsubscript{0.9}N is typically limited to 100~nm at \gls{RT} \cite{Evoy2000, Gonzalez2001}, the 500~nm thick Al\textsubscript{0.1}Ga\textsubscript{0.9}N spacer serves as an effective barrier which prevents beam-generated carriers from reaching the GaN buffer.
Consequently, the observed GaN emission in all the samples is attributed solely to the GaN \glspl{QW}.

To estimate the carrier density in the \glspl{QW} ($n_\text{QW}$), we first compute the lateral carrier generation rate distribution $F(r)$, where $r$ represents the lateral distance to the beam center using cylindrical coordinates. 
This is achieved by convolving the carrier generation rate distribution $F_0(r)$, deduced from the simulation, with a Gaussian distribution $G(r)$.
Specifically, $F_0(r)$ is obtained by summing the deposited energy over the $z$ direction, assuming that all carriers relax to the \gls{QW}.
$G(r)$ is characterized by a standard deviation $\sigma=22$~nm, which accounts for the broadening caused by carrier thermalization at 300~K \cite{Jahn2022}.
The resulting profile is normalized by its peak value at $r=0$ and fitted by a Voigt function, yielding a \gls{FWHM} of $\sim63$~nm (Fig.~\ref{Casino}\textbf{(c)}). 
It is important to note that our estimation does not consider the lateral carrier diffusion occurring in the surface barrier before carriers completely relax to the \gls{QW}. 
As a result, the calculated \gls{FWHM} underestimates the actual broadening of the carrier generation rate distribution in the \gls{QW}.
Meanwhile, the total generation rate, $G_\text{tot}$ (s\textsuperscript{-1}), of carriers in \gls{CL} can be estimated using the well-known equation \cite{Guthrey2020}:
\begin{equation}
	G_\text{tot} = \frac{I_\text{p}}{q} \cdot \frac{E_\text{dep}}{3 E_\text{g}} \hspace{1mm},
\end{equation}
where $I_\text{p}$ is the electron beam probe current, $q$ is the charge of an electron, $E_\text{dep}$ is the average energy deposited per electron in the sample, and $E_\text{g}$ is the bandgap of the sample. 
$E_\text{dep}$ is calculated as the difference between the beam energy ($E_\text{beam}$) and the energy lost through backscattered electrons ($E_\text{BSE}$).
In our case, $I_\text{p}=221$~to~233~pA, measured using a Faraday cup attached to the sample holder, $E_\text{g}=3.64$~eV for Al\textsubscript{0.1}Ga\textsubscript{0.9}N at 300~K \cite{Brunner1997}, $E_\text{beam}=5$~keV, and $E_\text{BSE}\approx1.14$~keV computed via Monte Carlo simulation.
Assuming that all generated carriers relax to the \gls{QW}, the carrier generation rate in the \gls{QW}, $G_\text{QW}$, is around $4.9$~to~$5.2 \cdot 10^{11}$~s\textsuperscript{-1}.
Under the assumptions of carrier lifetime being independent of carrier density and no carrier diffusion in the \gls{QW}, the maximal carrier density in the \gls{QW} at $r=0$ can be calculated as:
\begin{equation}
	n_{\text{QW}}^{\text{max}} = \frac{G_{\text{QW}} \cdot \tau}{2\pi \int_{0}^{\infty}F(r)r\mathrm{d}r} \hspace{1mm},
\end{equation}
with $\tau$ representing the carrier lifetime, which is typically around 1~ns for GaN/Al\textsubscript{0.1}Ga\textsubscript{0.9}N structures at \gls{RT} \cite{Podlipskas2019, Liu2019}.
Hence, the estimated maximum carrier density in the \glspl{QW} is approximately $8.2$~to~$8.7\cdot10^{12}$~cm\textsuperscript{-2}.
It should be noted that this value may be overestimated, considering the broader actual carrier distribution, as discussed previously.
Additionally, our assumption that all carriers relax to the \gls{QW} might not hold true in practice.
As a result, the average carrier density in the \glspl{QW} is expected to be on the order of $10^{12}$~cm\textsuperscript{-2}.

\section{3. Optical properties of surface GaN \glspl{QW}}
\label{QWs_sec}

\begin{figure}[b!]
	\centering
	\includegraphics[width=1\textwidth]{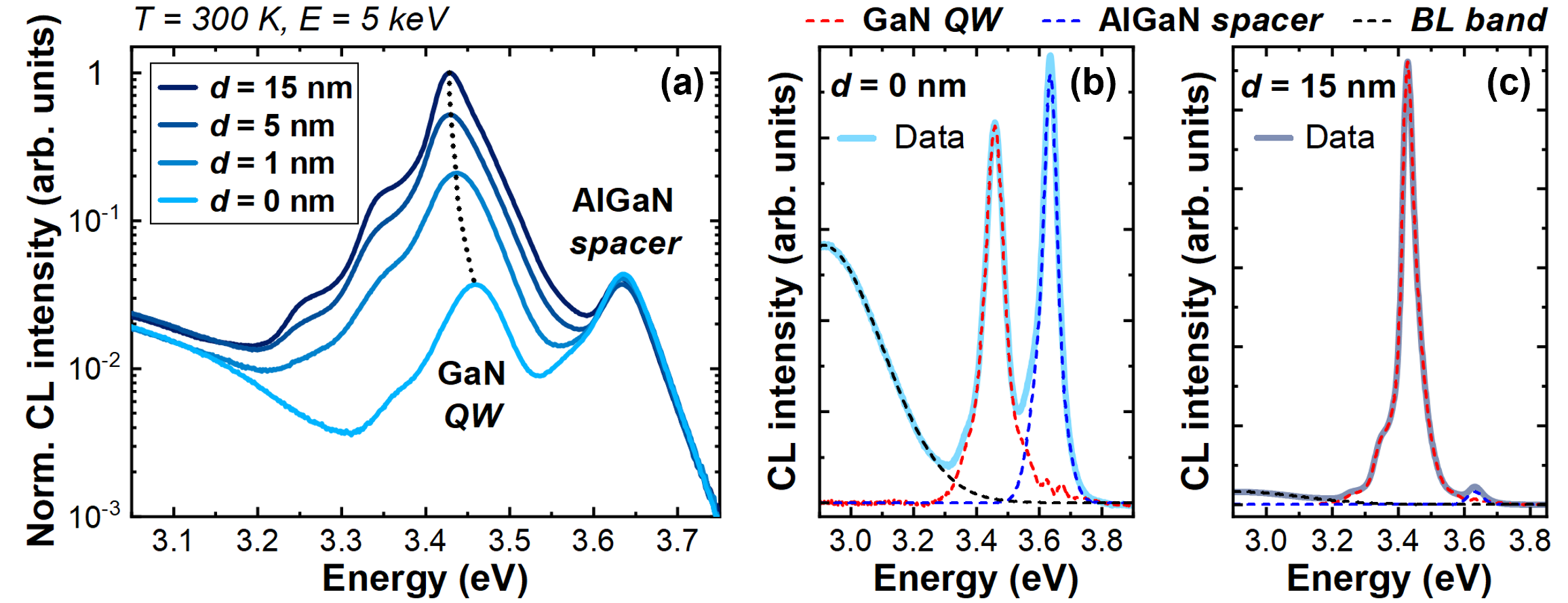}
	\caption{
		\textbf{(a)} \gls{CL} spectra of the surface GaN \glspl{QW}, excited by a 5~keV electron beam at 300~K. 
		The intensity scale is set to logarithmic to enhance the visibility of lower-intensity peaks.
		\textbf{(b, c)} For each \gls{CL} spectrum, a spectral deconvolution is applied to extract the peak energy and the integrated intensity of the GaN \gls{QW} and AlGaN spacer emissions.
	}
	\label{QWs-log}
\end{figure}

\hspace{\parindent}Since the \gls{QW} emission varies significantly with the surface barrier thickness, $d$, \gls{CL} spectra of the surface GaN \glspl{QW} are presented in logarithmic scale in Fig.~\ref{QWs-log}\textbf{(a)}, which improves the visibility of lower-intensity peaks. 
It is evident that, unlike surface GaAs \glspl{QW} \cite{Chang1993}, the emission intensity from our surface GaN \glspl{QW} remains considerably strong even in the absence of a surface barrier. 
In this logscale plot, the AlGaN spacer emission from all the samples exhibits almost identical intensity. 
This confirms that the injection level into the samples is nearly the same, unaffected by the varying~$d$.

For all the \gls{CL} spectra analyzed in this study, a spectral deconvolution is conducted to extract detailed information of the GaN \gls{QW} and AlGaN spacer emissions (Figs.~\ref{QWs-log}\textbf{(b,~c)}), in particular the peak energy and integrated intensity.	
The spectral deconvolution begins with a preliminary multi-peak fitting aimed at extracting the line shapes of the AlGaN emission and the background defect emission, \latin{i.e.}, the broad \gls{BL} band peaking at $\sim2.9$~eV \cite{Reshchikov2000}.
However, for the GaN \gls{QW} emission, which includes the \gls{ZPL} and multiple \gls{LO} phonon replicas, achieving an accurate fit is challenging. 
To address this, we deduced the \gls{QW} emission by subtracting the fitted \gls{BL} band and AlGaN emission from the raw data. 
This approach ensures a uniform treatment for all spectra, maintaining effective deconvolution of the overlapping \gls{QW} and AlGaN emissions (Fig.~\ref{QWs-log}\textbf{(b)}) while preserving the precise line shape of the entire \gls{QW} emission (Fig.~\ref{QWs-log}\textbf{(c)}). All the error bars in integrated intensity are estimated from the preliminary multi-peak fitting.

\begin{table}[b!]
	\centering
	\begin{tabular}{*{5}{c}}
		\toprule\toprule
		\textbf{\textit{d}} & \textbf{0 nm} & \textbf{1 nm} & \textbf{5 nm} & \textbf{15 nm}\\ \midrule
		GaN \gls{QW} & 3.459 eV & 3.437 eV & 3.429 eV & 3.429 eV \\
		AlGaN spacer & 3.635 eV & 3.635 eV & 3.635 eV & 3.633 eV \\
		\bottomrule
	\end{tabular}
	\caption{
		Peak energies of the GaN \gls{QW} and AlGaN spacer emissions deduced from the \gls{CL} spectra of surface GaN \glspl{QW} with varying surface barrier thickness ($d$), excited by a 5~keV electron beam at 300~K.
	}
	\label{E_peak}
\end{table}

\begin{figure}[t!]
	\centering
	\includegraphics[width=1\textwidth]{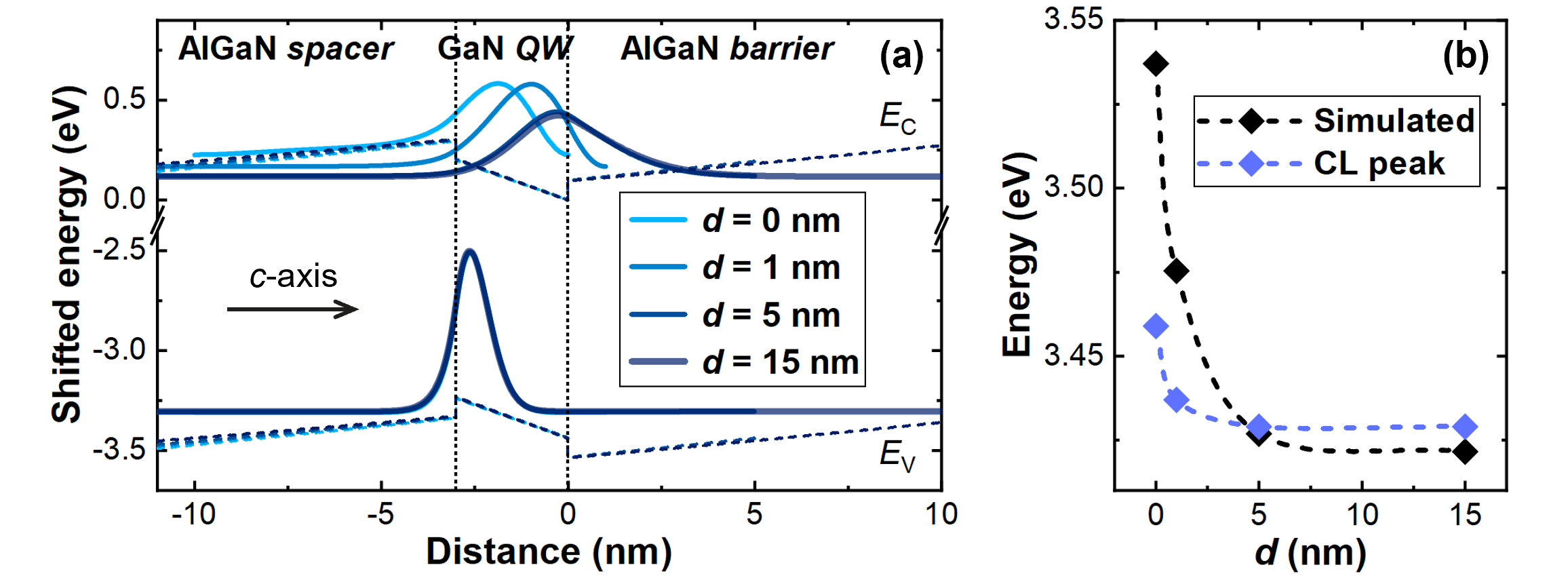}
	\caption{
		\textbf{(a)} Simulated band diagrams of surface GaN/AlGaN \glspl{QW} with varying surface barrier thickness ($d$), accompanied by the corresponding wavefunction probability densities of electrons and holes in the \gls{QW} region. 
		All the wavefunction probability density curves are shifted by the energy of their associated quantized state.
		$E_{\text{C}}$ and $E_{\text{V}}$ are the conduction band minimum and valence band maximum, respectively.
		Based on the simulations, 
		\textbf{(b)} the \gls{RT} interband transition energy is plotted as a function of $d$ (black diamonds), and compared to the \gls{CL} data (blue diamonds).
	}
	\label{BandStructures}
\end{figure}

The peak energies of the surface GaN \glspl{QW} are presented in Table~\ref{E_peak}. 
Notably, the peak energy of the AlGaN spacer emission remains nearly constant across all samples. 
In contrast, the GaN \gls{QW} peak exhibits a small blueshift for the sample with $d=1$~nm and an even larger blueshift for the uncapped \gls{QW} ($d=0$~nm).
To gain insights into these observations, we conducted band diagram calculations and analyzed the corresponding confined states using the commercial software, \textit{nextnano} \cite{nextnano}, as shown in Fig.~\ref{BandStructures}\textbf{(a)}.
For these polar surface \glspl{QW}, the presence of the free surface restricts the electron wavefunction penetration into the surface barrier, which leads to enhanced quantum confinement of carriers in the well when $d$ is small.
To further analyze the emission energy as a function of $d$, we compared the simulated interband transition energies with the GaN \gls{QW} peak energies extracted from \gls{CL} measurements (Fig.~\ref{BandStructures}\textbf{(b)}). 
While the overall $d$-dependent trends are in agreement, the simulated values for \glspl{QW} with $d=0,1$~nm are much higher than the experimental data. 
This disparity arises from the Dirichlet boundary condition applied at the free surface in the simulation, which does not consider the evanescent wave in the vacuum, thus leading to an overestimation of the confinement energy induced by the free surface.
Despite this discrepancy, the overall agreement between the two trends provides a reasonable explanation for the large blueshift observed in the uncapped \gls{QW}: carrier quantum confinement in this “well” is significantly enhanced by the free surface.

\begin{figure}[t!]
	\centering
	\includegraphics[width=1\textwidth]{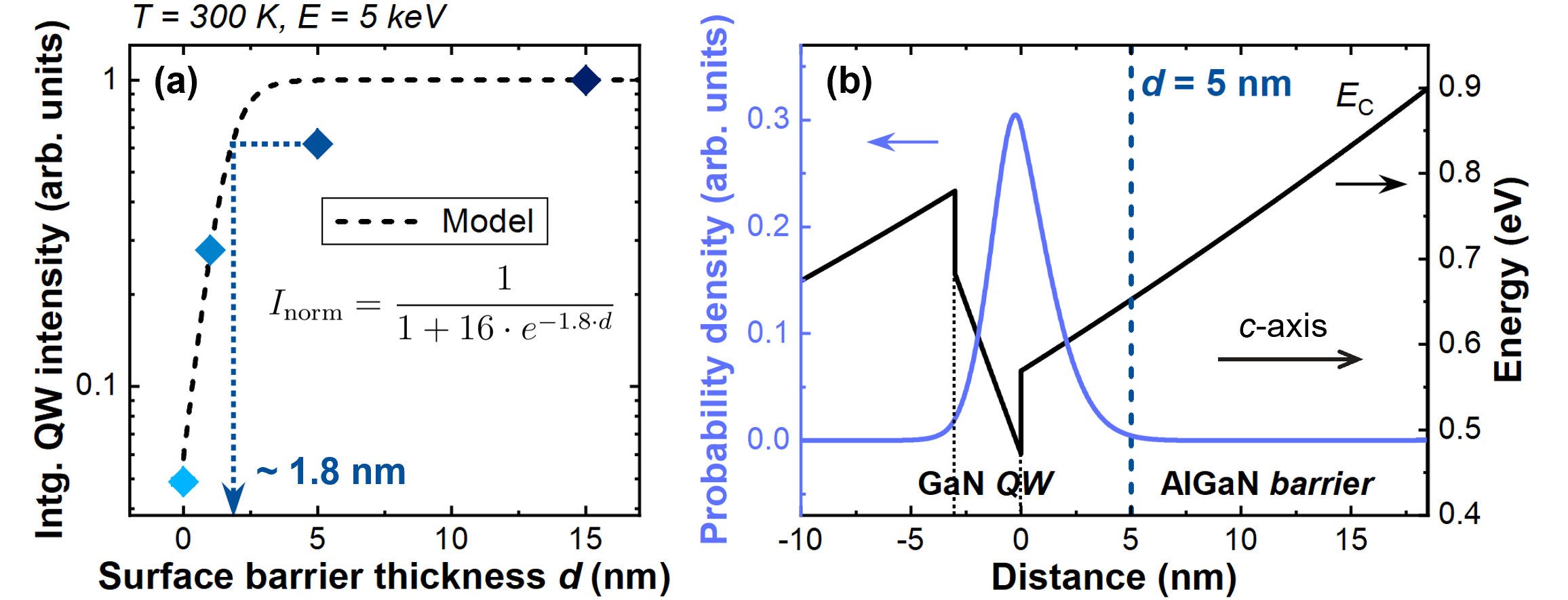}
	\caption{
		\textbf{(a)} $d$-dependent integrated \gls{QW} \gls{CL} intensity,  excited by a 5~keV electron beam at 300~K, compared to the model, which takes into account carrier tunneling from the well to the surface. 
		The intensity error bars are not visible in the plot as they are smaller than the size of the diamond symbol used.
		\textbf{(b)} Simulation of the electron wavefunction probability density across the barrier in the conduction band ($E_{\text{C}}$) imposed by a thick surface Al\textsubscript{0.1}Ga\textsubscript{0.9}N barrier.
		The dark blue dashed line corresponds to the position of the free surface of the surface GaN \gls{QW} with $d=5$~nm.
	}
	\label{Tunneling}
\end{figure}

In Fig.~\ref{Tunneling}\textbf{(a)}, the integrated \gls{QW} intensity is plotted against $d$, clearly demonstrating a nonlinear increase with increasing $d$.
However, this is contradictory to the simulation (Fig.~\ref{BandStructures}\textbf{(a)}): for larger $d$, the wavefunction overlap is smaller, thus the radiative recombination rate is lower. 
If the weight of non-radiative recombination was the same for all the samples, the \gls{QW} emission intensity should decrease with increasing $d$. 
Therefore, the observed increase in intensity is dominated by a reduction in non-radiative recombination channels.
Given the relation to $d$, these non-radiative channels are surface-related, \latin{i.e.}, \glspl{ST}. 

Since the physics of the envelope function formalism fails in vacuum as there is no longer any Bloch function, we will not use the simulated wavefunctions for the comparison of the \gls{QW} emission intensities.
To model the $d$-dependent \gls{QW} emission intensity, we take into account the probability of carrier tunneling from the well to the surface \cite{Chang1993}, where carriers could potentially become trapped without emitting light, \latin{i.e.}, recombine non-radiatively.
First of all, the measured \gls{QW} emission intensity is correlated to the internal quantum efficiency of the sample, which is the ratio between the \gls{RRR} and the \gls{ERR}. 
Given that these samples possess similar structures and were grown under identical conditions, and that their difference in confinement has a negligible impact on the emission, as previously discussed, it is reasonable to assume that \gls{RRR} in these \glspl{QW} remains constant.
Subsequently, the \gls{QW} intensity should be inversely proportional to \gls{ERR}:
\begin{equation}
	\label{I_QW}
	I_{\text{QW}} \propto 1/R_{\text{eff}} \hspace{1mm}, \qquad \text{with} \quad R_{\text{eff}} = R_0 + R_{\text{NR, ST}} \hspace{1mm}, 
\end{equation}
where $R_0$ represents the effective recombination rate in the absence of \glspl{ST}, \latin{i.e.}, \glspl{QW} with very large $d$. 
$R_{\text{NR, ST}}$ denotes the non-radiative recombination rate associated with \glspl{ST}, which can be expressed as:
\begin{equation}
	\label{R_NR,ST}
	R_{\text{NR, ST}} \propto P_{\text{ST}} \cdot \int_{d}^{\infty} \Psi^2(x) \mathrm{d}x \hspace{1mm},
\end{equation}   	
where $P_{\text{ST}}$ is the probability of a surface carrier being non-radiatively trapped by \glspl{ST}, and $\Psi$ is the carrier Schrödinger wavefunction, with $x$ the distance measured along the $c$-axis. 
Here, $x=0$ is defined at the \gls{QW}-barrier interface. 
Hence, the integral of the probability density, $\Psi^2$, from $d$ to infinity, is associated with the probability of carrier tunneling from the well to the surface.
In our case, due to the presence of the built-in field, electrons in the well are more likely to cross the barrier (Fig.~\ref{BandStructures}\textbf{(a)}).
Therefore, they are regarded as the main carriers responsible for the tunneling from the well to the surface.
Assuming a similar, albeit minor, surface band bending across all samples, the barrier height for electrons in the \glspl{QW} is approximately the conduction band offset, $\Delta E_{\text{C}}$, between the GaN \gls{QW} and Al\textsubscript{0.1}Ga\textsubscript{0.9}N barrier. 
This value can be estimated by considering the bandgap ($E_{\text{g}}$) at 300~K of GaN ($3.42$~eV \cite{Brunner1997}) and Al\textsubscript{0.1}Ga\textsubscript{0.9}N ($3.64$~eV \cite{Brunner1997}), along with the “common anion rule” \cite{Rosencher2002}: 
$\Delta E_{\text{C}} \approx 0.7 \Delta E_{\text{g}}=154$~meV.
The one-dimensional Schrödinger equation for an electron in the barrier can be written in the form:
\begin{equation}
	\label{Schrodinger}
	\frac{d^2}{dx^2} \Psi(x) = \frac{2m^*}{\hbar^2} \Delta E_{\text{C}} \Psi(x) = {\kappa}^2 \Psi(x) \hspace{1mm}, \qquad \text{where} \quad {\kappa}^2 = \frac{2m^*}{\hbar^2} \Delta E_{\text{C}} \hspace{1mm}.
\end{equation}
Here $\hbar$ is the reduced Planck's constant and $m^*$ is the effective mass of the electron in Al\textsubscript{0.1}Ga\textsubscript{0.9}N ($m^* \approx 0.2 m_0$ \cite{Vurgaftman2003}, with $m_0$ the electron rest mass). 
The solution of the equation is in the form of an evanescent wave: $\Psi(x) = \Psi_0 e^{-\kappa x}$, with $\Psi_0$ a constant coefficient. By inserting this expression into Eq.~\ref{R_NR,ST}, we obtain:
\begin{equation}
	\label{d-R_NR,ST}
	R_{\text{NR, ST}} \propto P_{\text{ST}} \cdot e^{-2 \kappa d}  \hspace{1mm}.
\end{equation}
Given that all the samples exhibit comparable material quality, we can treat $R_0$ and $P_{\text{ST}}$ as constants independent of the variable $d$. 
Subsequently, by combining Eqs.~\ref{I_QW} and \ref{d-R_NR,ST}, we can model the $d$-dependent normalized \gls{QW} intensity using the following formula:
\begin{equation}
	I_{\text{norm}}(d) = \frac{1}{1 + C \cdot e^{-2\kappa d}}  \hspace{1mm}, \qquad \text{where} \quad \lim_{d \to \infty} I_{\text{norm}} \to 1 \hspace{1mm}.
\end{equation}
Here $\kappa\approx0.9$, calculated using Eq.~\ref{Schrodinger}, and $C$ is a parameter that can be deduced from the data.
The \gls{QW} \gls{CL} intensity in Fig.~\ref{Tunneling}\textbf{(a)} is normalized by the value at $d=15$~nm, which aligns with the model, given that the term $e^{-2\kappa d}$ is $\sim 10^{-12}$ at $d=15$~nm, using the previously calculated value of $\kappa$.
As shown in Fig.~\ref{Tunneling}\textbf{(a)}, our model (black dashed curve) successfully captures most of the experimental data, except for the point at $d=5$~nm.

Indeed, if we compute the electron wavefunction probability density in a well with large $d$ (Fig.~\ref{Tunneling}\textbf{(b)}), the penetration is no longer affected by the boundary condition at the free surface.
Electrons confined in the well have a probability of less than 1$\%$ to tunnel through a 5~nm Al\textsubscript{0.1}Ga\textsubscript{0.9}N barrier (indicated by the dark blue dashed line).
The observed increase in intensity when $d$ increases from 5~nm to 15~nm suggests that the \gls{d_eff} of the 5~nm thick AlGaN barrier, acting as an “electron blocking layer”, is significantly smaller than its physical size.
In fact, for the same intensity in the model's curve, the corresponding barrier thickness is $\sim1.8$~nm (indicated by the dark blue dotted arrow in Fig.~\ref{Tunneling}\textbf{(a)}), \latin{i.e.}, $\sim3.2$~nm less than the actual thickness.
A possible explanation for this discrepancy is the presence of percolative paths in the AlGaN barrier caused by alloy disorder \cite{Nath2013}.
The random distribution of Al atoms in the barrier results in regions of varying Al content, known as alloy disorder. 
In other words, if we consider the same Al content in the barrier, it implies that \gls{d_eff} is locally changing.
Since $d=5$~nm is at the limit of carrier tunneling (Fig.~\ref{Tunneling}\textbf{(b)}), for regions where \gls{d_eff}~$>5$~nm, there is no significant impact on the emission. 
However, in regions where \gls{d_eff}~$<5$~nm, the intensity can be dramatically affected.
Thus, the overall consequence of the percolative paths within the 5~nm thick AlGaN barrier is a reduction in \gls{d_eff}. 
Considering the typical scale of alloy disorder in the III-nitride system, \latin{i.e.}, the average size of the Al-free regions in our case, is $\sim3$~nm \cite{Li2017}, the disorder-induced percolation effect is consistent with the observed difference between the physical and \acrlong{d_eff} for $d=5$~nm.
Regarding the other two samples with smaller $d$, fluctuations in both directions (decreasing and increasing $d$) lead to changes in the intensity. 
Therefore, the order of magnitude of the intensity for these samples is less affected by the percolation effect. 
As a result, unlike the sample with $d=5$~nm, the values of these samples align well with the model, as shown in Fig.~\ref{Tunneling}\textbf{(a)}.

To summarize the findings discussed so far, we model the $d$-dependent \gls{QW} intensity by considering the carrier tunneling effect, and deduce that the presence of alloy disorder in the AlGaN barrier reduces its \acrlong{d_eff}.
This leads to surface-sensitive \gls{QW} emission even with a relatively thick barrier. 
Among all the samples studied, only the \gls{QW} with $d=15$~nm appears to be entirely unaffected by surface effects.
To further understand the impact of alloy disorder in the barrier, additional investigations, such as transmission electron microscopy, and comparison with \glspl{QW} using GaN barriers, which lack alloy disorder, are necessary. 
These investigations are currently in progress in our group.

\begin{figure}[h!]
	\centering
	\includegraphics[width=1\textwidth]{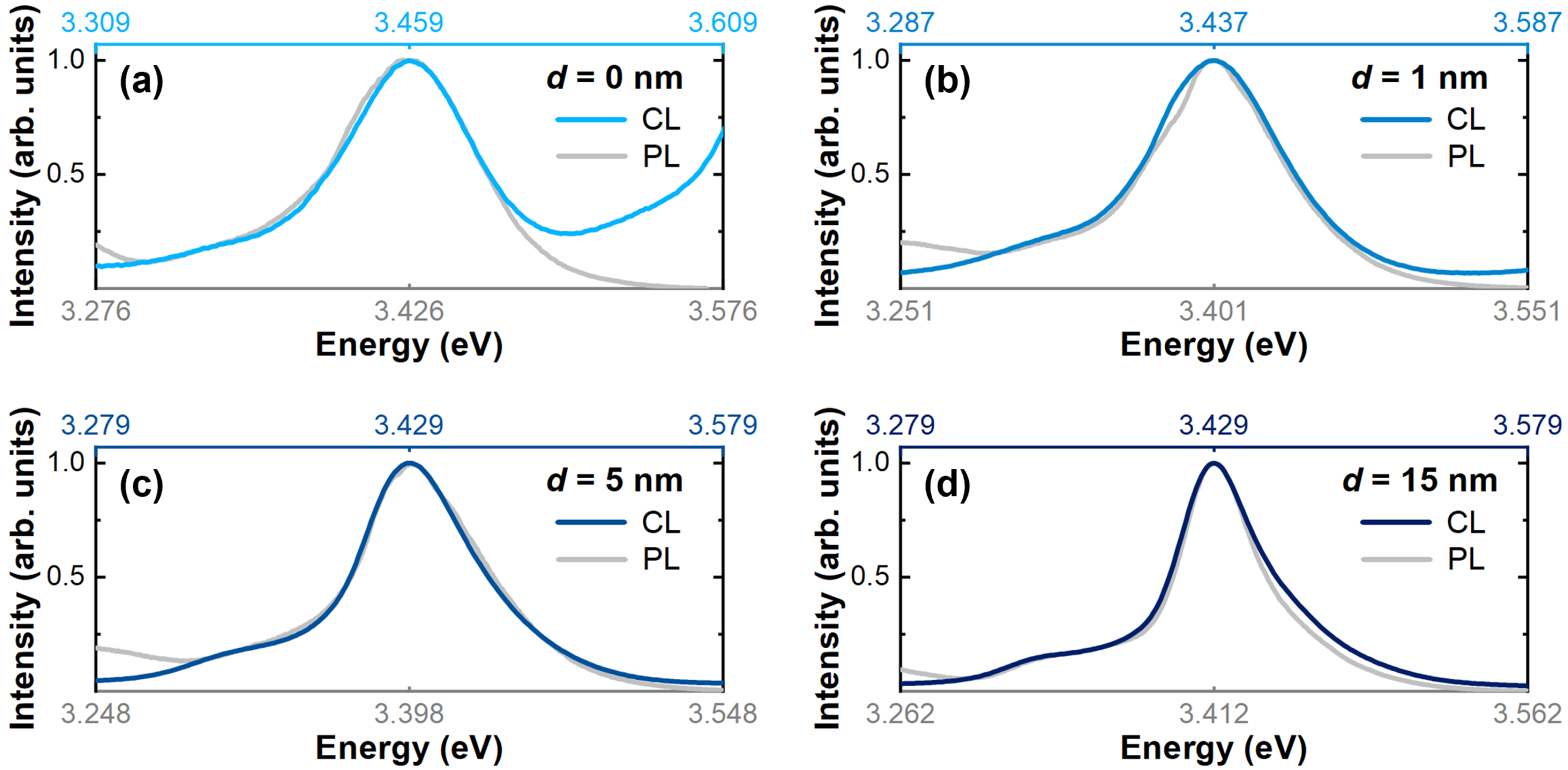}
	\caption{
		Comparison of the \gls{QW} emission line shape under different injection conditions: lower carrier density ($\sim10^{10}$~cm\textsuperscript{-2}) in \gls{PL} and higher carrier density ($\sim10^{12}$~cm\textsuperscript{-2}) in \gls{CL} at 300~K, for surface GaN \glspl{QW} with varying barrier thickness:
		\textbf{(a)} $d=0$~nm, \textbf{(b)} $d=1$~nm, \textbf{(c)} $d=5$~nm, and \textbf{(d)} $d=15$~nm.
		The energy axis of the \gls{CL} spectra is shifted to match the peak energy in the \gls{PL} spectra while maintaining the same scale.
	}
	\label{PLvsCL}
\end{figure}

In order to investigate the change in the \gls{QW} emission line shape under different injection levels, we compare the \gls{CL} spectra with the \gls{PL} spectra of the same samples.
The \gls{PL} spectra were obtained using a \gls{cw} laser with a power density of $\sim16$~W/cm\textsuperscript{2} at a wavelength of 325~nm. 
Under this excitation condition, assuming that nearly all carriers relax to the \gls{QW} and using the previously employed carrier lifetime of 1~ns, the estimated carrier density in the \glspl{QW} is on the order of $10^{10}$~cm\textsuperscript{-2} in \gls{PL} \cite{Trivino2015}, which is nearly two orders of magnitude lower than the estimated carrier density in \gls{CL} (SI Sec.~\hyperref[CASINO_sec]{2}).
As depicted in Fig.~\ref{PLvsCL}, the \gls{CL} peaks are generally blueshifted by $\sim30$~meV, which can be attributed to a stronger screening of the built-in field in the \gls{QW} caused by the higher carrier density.
On the other hand, the line shape of the \gls{QW} peaks remains nearly identical between \gls{CL} and \gls{PL} spectra.
If the carrier density in \gls{CL} were above the critical density of the Mott transition, a much more asymmetric line shape would be expected due to band filling of the continuum states \cite{Rossbach2014}.
Our observation strongly suggests that in the \gls{CL} measurements, carriers in the \glspl{QW} predominantly exist as excitons rather than unbound electron-hole pairs. 
Consequently, the emission studied in this work is primarily governed by excitonic features.

\section{4. Thickness determination of MoS\textsubscript{2}}
\label{MoS2_thickness_sec}

\begin{figure}[h!]
	\centering
	\includegraphics[width=1\textwidth]{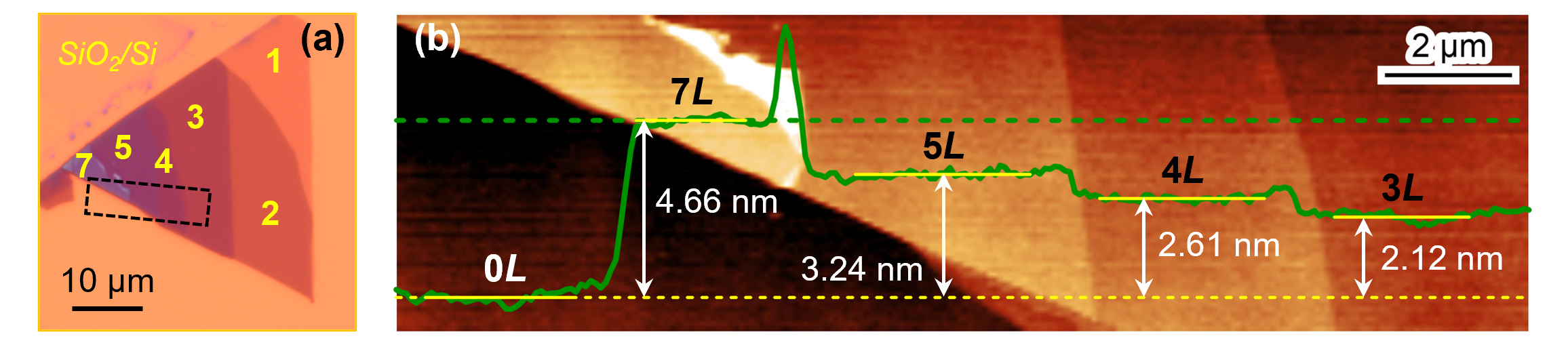}
	\caption{
		\textbf{(a)} Optical micrograph of the selected MoS\textsubscript{2} flake, deposited on a SiO\textsubscript{2}/Si substrate. 
		The numbers in yellow indicate the number of MoS\textsubscript{2} \glspl{1L} in the corresponding region, determined by \gls{AFM} and Raman spectroscopy.
		\textbf{(b)} \gls{AFM} height image of the rectangular area outlined by the black dashed line in \textbf{(a)}. 
		The thickness of each layer is determined from the height profile (green solid curve) taken along the green dashed line in the \gls{AFM} image.
		The “$nL$” labels indicate that the MoS\textsubscript{2} thickness in the corresponding region is $n$~\glspl{1L}.
	}
	\label{AFM}
\end{figure}

\hspace{\parindent}\Gls{2D} MoS\textsubscript{2} flakes were first obtained by mechanically exfoliating bulk MoS\textsubscript{2} crystals, and deposited onto a Si substrate pre-coated with a 275~nm thick SiO\textsubscript{2} layer.
This particular thickness of oxide was chosen to optimize the visibility of \gls{1L}-MoS\textsubscript{2} under the optical microscope, based on light interference \cite{Benameur2011}.
The flakes of interest were initially identified using optical microscopy (Fig.~\ref{AFM}\textbf{(a)}), and their thickness was subsequently determined by \gls{AFM} (Fig.~\ref{AFM}\textbf{(b)}).
For our MoS\textsubscript{2} samples with lateral thickness variation, we estimated the layer thickness in different regions based on the typical \gls{1L}-MoS\textsubscript{2} thickness of $\sim0.65$~nm. 
However, it must be acknowledged that in the \gls{AFM} measurements, \gls{1L}-flakes on bare substrates showed a broad distribution in heights, ranging from 0.6~to~0.9~nm \cite{Lee2010}, which may be due to the presence of adsorbates beneath the flake or other flake-substrate interactions \cite{Nemes-Incze2008}.

\begin{figure}[h!]
	\centering
	\includegraphics[width=1\textwidth]{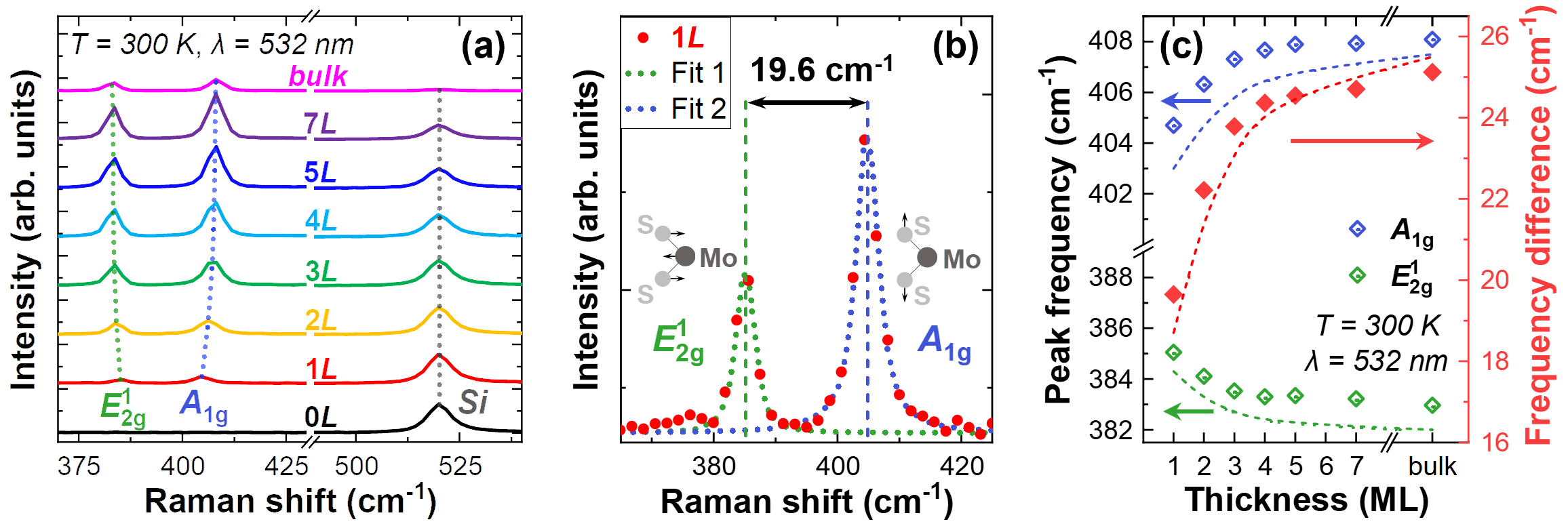}
	\caption{
		\textbf{(a)} Raman spectra obtained from various regions of the MoS\textsubscript{2} flake as illustrated in Fig.~\ref{AFM}\textbf{(a)}. 
		Each spectrum is labelled with the layer thickness (number of \glspl{1L}, n$L$) estimated through \gls{AFM} measurements.
		A double-Voigt fit is applied to each spectrum to extract the peak frequencies of the $E_{\text{2g}}^1$ and $A_{\text{1g}}$ Raman modes.
		\textbf{(b)} In the \gls{1L}-MoS\textsubscript{2} region ($1L$), the difference between the two peaks is $\sim19.6$~cm\textsuperscript{-1}, which falls within the expected range of \gls{1L}-MoS\textsubscript{2}. 
		\textbf{(c)} The frequencies of the $E_{\text{2g}}^1$ and $A_{\text{1g}}$ Raman modes (left vertical axis) and their difference (right vertical axis) are plotted as a function of layer thickness.
		Our results (diamonds) are compared with the expected trends (dashed curves) \cite{Lee2010}.
	}
	\label{Raman}
\end{figure}

To double-check the layer thickness, Raman spectroscopy measurements were carried out in different regions using a \gls{cw} 532~nm laser in an air ambient environment (Fig.~\ref{Raman}\textbf{(a)}).
This method is based on the fact that the Raman frequencies of the $E_{\text{2g}}^1$ mode (in-plane opposite vibrations of S and Mo atoms, illustrated in Fig.~\ref{Raman}\textbf{(b)}) and the $A_{\text{1g}}$ mode (out-of-plane vibration of S atoms in opposite directions, illustrated in Fig.~\ref{Raman}\textbf{(b)}) are highly sensitive to MoS\textsubscript{2} thickness within the range of 1-4~\glspl{1L} \cite{Lee2010}. 
Especially, the difference between the two peaks in \gls{1L}-MoS\textsubscript{2} generally falls within the range of 18-21~cm\textsuperscript{-1}, regardless of the laser \cite{Li2012} or substrate \cite{Buscema2014} used.
Therefore, this feature is commonly used to identify \gls{1L}-MoS\textsubscript{2}.
In our case, the observed thickness-dependent Raman peaks and the frequency difference between the two peaks (diamonds in Fig.~\ref{Raman}\textbf{(c)}) align well with the reported trend (dashed curves in Fig.~\ref{Raman}\textbf{(c)}) \cite{Lee2010}, thus corroborating the thickness determined by \gls{AFM}.
Furthermore, the presence of a \gls{1L}-MoS\textsubscript{2} region ($1L$) in the flakes of interest, which is crucial for our study, is well confirmed (Fig.~\ref{Raman}\textbf{(b)}).

\section{5. \gls{CL} data processing}
\label{CL_processing_sec}

\begin{figure}[b!]
	\centering
	\includegraphics[width=1\textwidth]{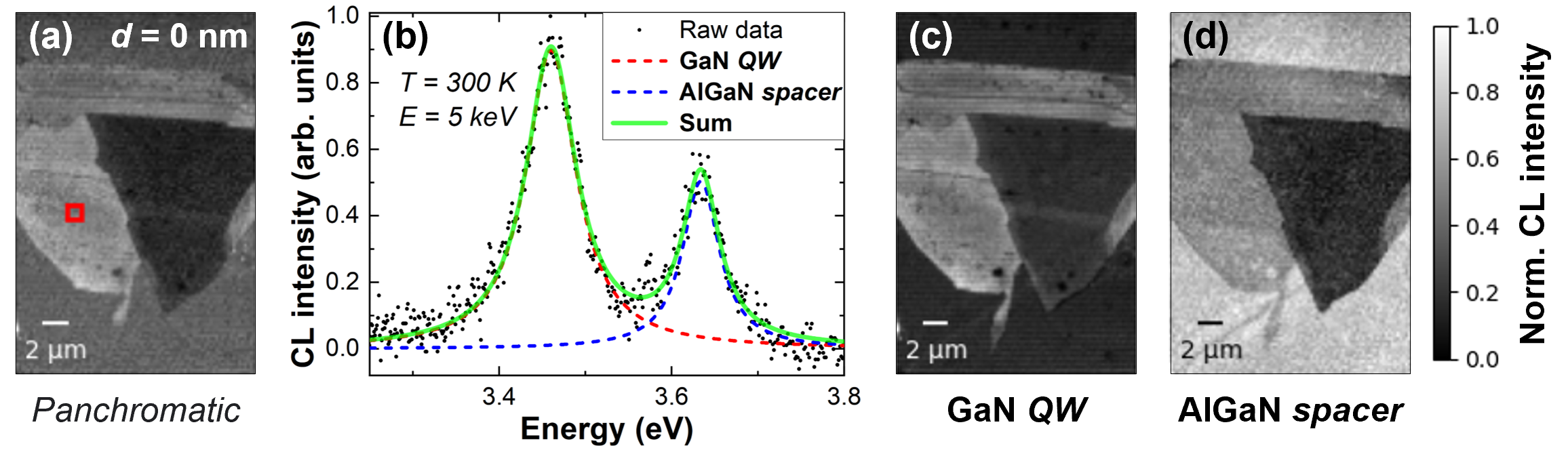}
	\caption{
		\textbf{(a)} Panchromatic \gls{CL} map of the uncapped QW ($d=0$~nm) coated with MoS\textsubscript{2}, measured at 300~K using a 5~keV electron beam.
		\textbf{(b)} Normalized \gls{RT} \gls{CL} spectrum acquired at the pixel represented by the red square in \textbf{(a)}. 
		By performing multi-peak fitting for all pixels, integrated intensity maps of the \textbf{(c)} GaN \gls{QW} and \textbf{(d)} AlGaN spacer emissions are generated.
		All maps are individually normalized by their maximum and minimum values.	
	}
	\label{CL-maps}
\end{figure}

\hspace{\parindent}The hyperspectral analysis was conducted using the \textit{hyperspy} package in Python \cite{francisco_de_la_pena_2022_7263263}. 
For each pixel in the hyperspectral map, a multi-peak fitting is employed to extract the integrated intensity of the GaN \gls{QW} and AlGaN spacer emissions (Figs.~\ref{CL-maps}\textbf{(a,~b)}), which enables the generation of integrated intensity maps for each emission within the scanned area (Figs.~\ref{CL-maps}\textbf{(c,~d)}).

\begin{figure}[h!]
	\centering
	\includegraphics[width=1\textwidth]{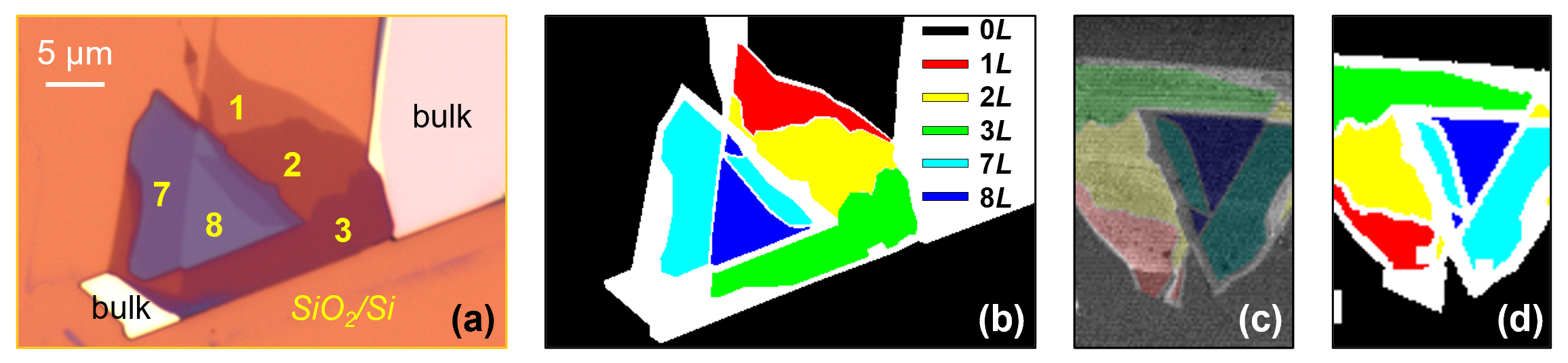}
	\caption{
		\textbf{(a)} High-resolution optical micrograph of the MoS\textsubscript{2} flake deposited on a SiO\textsubscript{2}/Si substrate, which is 
		\textbf{(b)} segmented according to thickness-sensitive color contrast.
		\textbf{(c)} The segmented image is rotated and resized to match the flake in the \gls{CL} map. 
		\textbf{(d)} After cropping, pixel binning, and dead pixel identification, a mask of the \gls{CL} map is generated
	}
	\label{CL-mask}
\end{figure}

To extract information from regions with different MoS\textsubscript{2} thicknesses, we employed image segmentation on the high-resolution optical micrograph of the flake (Figs.~\ref{CL-mask}\textbf{(a,~b)}).
This is based on the thickness-sensitive color contrast mentioned earlier.
Consequently, regions with different thicknesses are assigned specific colors, \latin{i.e.}, defined pixel values.
The resulting segmented image was then rotated and resized to align it with the contour of the flake in the panchromatic \gls{CL} map (Fig.~\ref{CL-mask}\textbf{(c)}). 
The reshaped image was subsequently cropped and binned to generate a mask for the \gls{CL} map, ensuring that its image size and number of pixels are the same as the \gls{CL} map (Fig.~\ref{CL-mask}\textbf{(d)}).
Due to the lower resolution of the \gls{CL} map, pixels located at the boundaries of the segmented regions exhibit intermediate values after the binning process.
To address this, all these new pixel values were converted to (255, 255, 255) (white in color), which represent “dead pixels” and are excluded from further data processing and analysis.
For all the \gls{CL} maps in this study, we generated an associated mask where the color of each pixel represents the number of MoS\textsubscript{2} \glspl{1L} at that position.

The use of thickness-related masks is crucial to our study. 
Firstly, these masks enable the calculation of the average background intensity ($I_{\text{Bkg}}$) in different emission maps, which allows equivalent normalization of the maps and facilitates comparison across different samples.
Specifically, the normalized intensity is calculated using $I_{\text{norm}} = I /I_{\text{Bkg}}$, so that the average background intensity is set to 1 for all maps. 
In this way, the contrast observed in each map represents the relative change in intensity induced by the presence of MoS\textsubscript{2}, allowing direct comparisons between different maps.
Secondly, instead of comparing solely the intensity maps that primarily reflect spatial variations related to the surface morphology of the sample, we also focus on the spectral domain by comparing the average spectra obtained from regions coated with MoS\textsubscript{2} of different thicknesses.
The advantage of this analysis is twofold.
On the one hand, the average spectrum captures the general properties of the interface effect associated with MoS\textsubscript{2} thickness and effectively minimizes the impact of microscale fluctuations arising from surface roughness or other factors that are not the main focus of this study.
On the other hand, by averaging a large number of spectra, the resulting spectrum has a higher signal-to-noise ratio. 
This facilitates quantitative analyses, including the peak identification (\latin{e.g.}, \gls{ZPL} and \gls{LO} phonon replicas of the GaN \gls{QW} emission) and the calculation of energy-dependent intensity contrasts (see Fig.~\ref{15nm}\textbf{(b)} in the main text).
Overall, the use of thickness-related masks enables equivalent data normalization, spectral analysis, and quantitative calculations, which improves the interpretability and reliability of our results.

\section{6. MoS\textsubscript{2} on bulk GaN epilayer}
\label{MoS2_on_buffer_sec}

\begin{figure}[b!]
	\centering
	\includegraphics[width=1\textwidth]{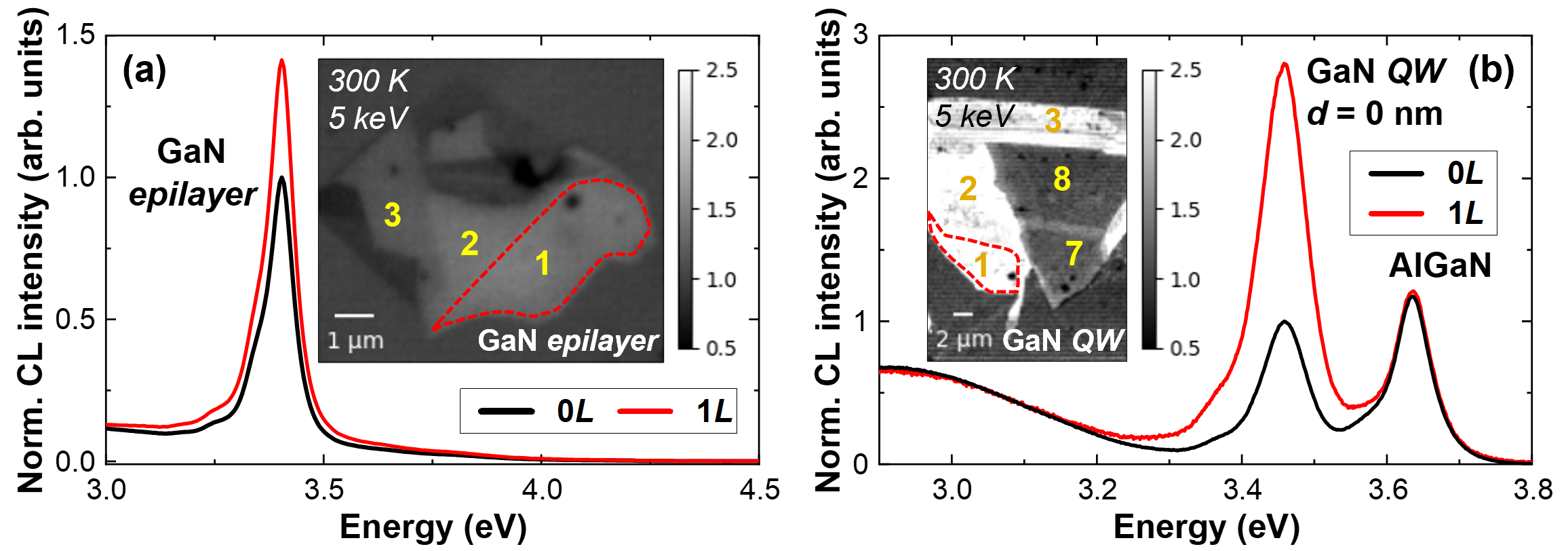}
	\caption{
		Average \gls{RT} \gls{CL} spectra of the background emission (in absence of MoS\textsubscript{2}, $0L$) and the emission from the region covered by \gls{1L}-MoS\textsubscript{2} ($1L$) extracted from 
		\textbf{(a)} the GaN epilayer and 
		\textbf{(b)} the uncapped GaN \gls{QW} ($d=0$~nm), 
		both coated with a MoS\textsubscript{2} flake.
		The corresponding integrated GaN \gls{CL} intensity maps are shown in the insets, with the $1L$ region indicated by the red dashed line. 
		These maps are normalized by their average background intensity, respectively.
	}
	\label{MoS2onGaN}
\end{figure}

\hspace{\parindent}To confirm the hypothesis of the surface passivation effect, we deposited an MoS\textsubscript{2} flake of 1-3~\glspl{1L} on a bulk GaN  epilayer, identical to the GaN buffer used to grow the \glspl{QW} (SI Sec.~\hyperref[methods_sec]{1}).
\gls{CL} measurement was performed on this flake at 300~K using 5~keV excitation. 
The resulting GaN intensity map is normalized (inset of Fig.~\ref{MoS2onGaN}\textbf{(a)}) and compared to the normalized GaN \gls{QW} intensity map of the uncapped GaN \gls{QW} ($d=0$~nm) coated with \gls{2D} MoS\textsubscript{2} (inset of Fig.~\ref{MoS2onGaN}\textbf{(b)}). 
Through normalization, both maps have the same average background intensity, \latin{i.e.}, 1. 
Upon comparison, it is evident that both maps demonstrate enhanced GaN emission in the presence of MoS\textsubscript{2}.
However, the enhancement is obviously stronger for the uncapped GaN \gls{QW} emission than for the GaN epilayer emission.
To quantitatively visualize this difference, average \gls{CL} spectra of the bare region ($0L$) and the region coated with \gls{1L}-MoS\textsubscript{2} ($1L$) are compared.
For the GaN epilayer, the emission is enhanced by \gls{1L}-MoS\textsubscript{2}, with a peak intensity ratio of $\sim1.4$ (Fig.~\ref{MoS2onGaN}\textbf{(a)}).
In contrast, for the uncapped \gls{QW}, only the GaN \gls{QW} emission is enhanced, but with a higher peak intensity ratio of $\sim2.8$ (Fig.~\ref{MoS2onGaN}\textbf{(b)}).
This discrepancy in MoS\textsubscript{2}-induced intensity enhancement between the two cases is consistent with the surface passivation effect.
In the uncapped \gls{QW}, the detected GaN emission comes solely from the surface region, whereas in the GaN epilayer, the emission arises from both surface and bulk regions.
As a result, the enhancement is much greater in the former case due to the direct influence of surface passivation.

\section{7. MoS\textsubscript{2} on GaN \gls{QW}, $\boldsymbol{d=15}$~nm}
\label{15nm_sec}

\glsreset{ST}
\glsreset{DoS}
\begin{figure}[b!]
	\centering
	\includegraphics[width=1\textwidth]{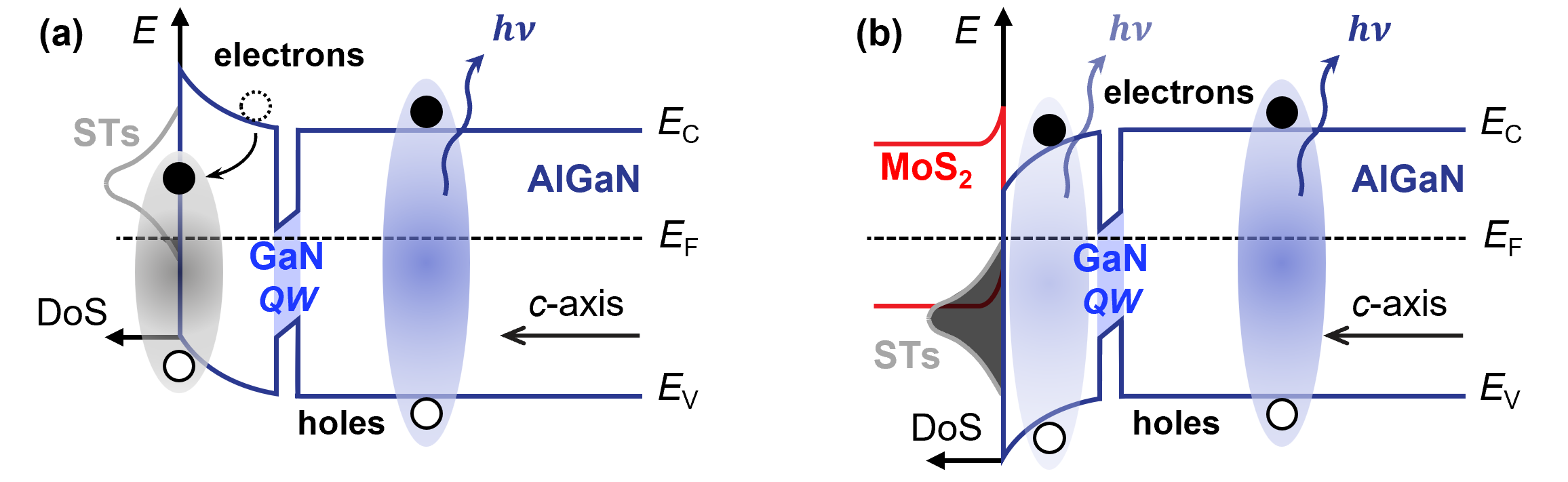}
	\caption{
		Surface band bending of \textbf{(a)} bare GaN \gls{QW} with $d=15$~nm  and \textbf{(b)} the \gls{QW} coated with MoS\textsubscript{2} following interfacial charge transfer \cite{Jain2020}.
		The \gls{DoS} of \glspl{ST} in the AlGaN surface barrier is depicted for both scenarios, with the black shaded part representing occupied states.
		The grey ellipse represents an exciton captured by a \gls{ST} and does not emit light, while the blue ellipses represent free excitons that can emit light through radiative recombination.
	}
	\label{MoS2on15nm-Pass}
\end{figure}

\hspace{\parindent}The presence of \gls{1L}-MoS\textsubscript{2} enhances the AlGaN emission from the GaN \gls{QW} with $d=15$~nm (see Fig.~\ref{15nm}\textbf{(b)} in the main text). 
This enhancement can be attributed to the surface passivation induced by the deposition of \gls{1L}-MoS\textsubscript{2}, as illustrated in Fig.~\ref{MoS2on15nm-Pass}.
In the case of a bare \gls{QW} without the deposition of MoS\textsubscript{2} (Fig.~\ref{MoS2on15nm-Pass}\textbf{(a)}), potential emission in the AlGaN surface barrier is quenched due to the presence of \glspl{ST}.
Consequently, the AlGaN emission primarily originates from the spacer region, situated far from the surface.
Upon the deposition of \gls{1L}-MoS\textsubscript{2} (Fig.~\ref{MoS2on15nm-Pass}\textbf{(b)}), the change in surface band bending \cite{Jain2020} results in the filling of \glspl{ST} by charges transferred through the \gls{vdW} interface.
As the occupied \glspl{ST} are no longer capable of capturing excitons in the surface region, the emission from the surface AlGaN barrier could potentially be enhanced, which results in an overall increase in the AlGaN emission in the presence of \gls{1L}-MoS\textsubscript{2} coating.

\section{8. MoS\textsubscript{2} on GaN \glspl{QW}, $\boldsymbol{d=1}$~nm and $\boldsymbol{d=5}$~nm}
\label{1nm_5nm_sec}

\begin{figure}[h!]
	\centering
	\includegraphics[width=1\textwidth]{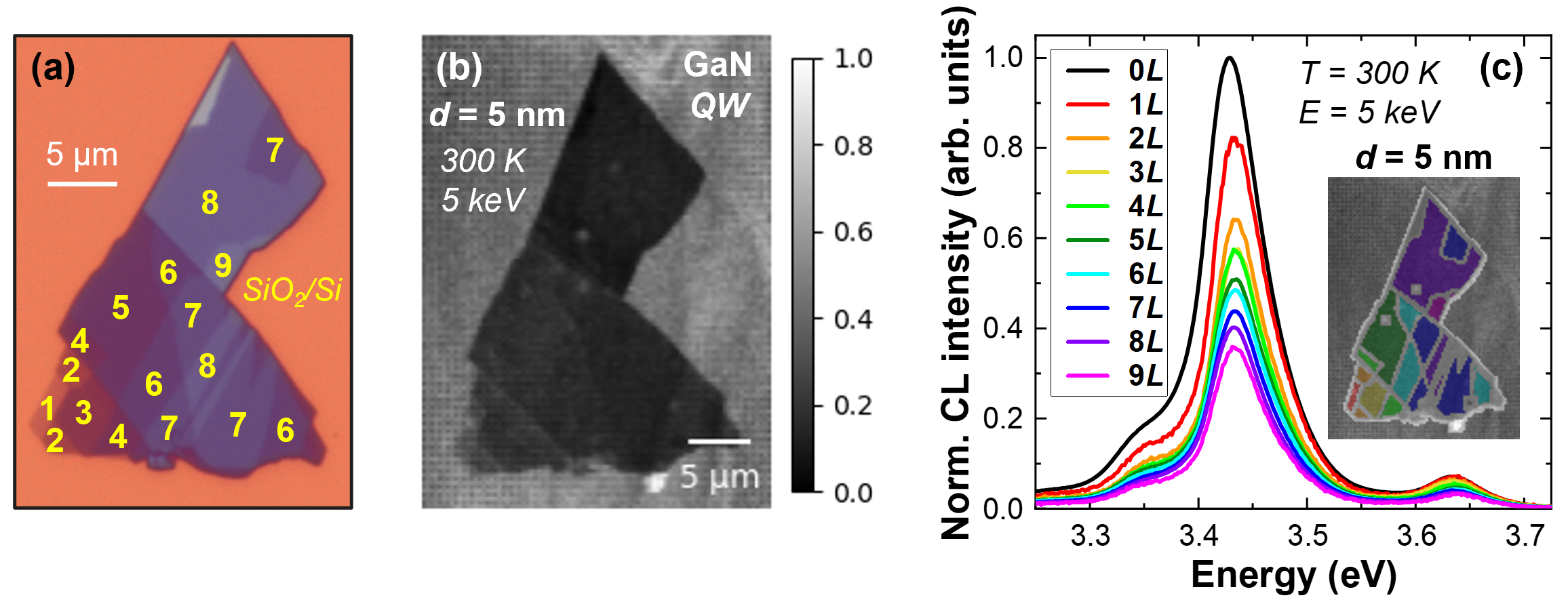}
	\caption{
		\textbf{(a)} Optical micrograph of the MoS\textsubscript{2} flake on a SiO\textsubscript{2}/Si substrate.	
		The yellow numbers indicate the number of MoS\textsubscript{2} \glspl{1L} in the corresponding region.
		\textbf{(b)} Normalized integrated \gls{CL} intensity map of the GaN \gls{QW} emission obtained from the MoS\textsubscript{2}-coated \gls{QW} with $d=5$~nm, excited by a 5~keV electron beam at 300~K.
		The normalization is done with respect to the maximum and minimum values in the map.
		\textbf{(c)} Average \gls{RT} \gls{CL} spectra extracted from regions with different MoS\textsubscript{2} thicknesses. 
		The corresponding regions are shown in the inset. 
	}
	\label{MoS2on5nm}
\end{figure}

\hspace{\parindent}Detailed \gls{CL} results of the \gls{QW} with $d=5$~nm are presented in Fig.~\ref{MoS2on5nm}.
This includes the optical micrograph of the MoS\textsubscript{2} flake deposited on a SiO\textsubscript{2}/Si substrate (Fig.~\ref{MoS2on5nm}\textbf{(a)}), the integrated \gls{QW} \gls{CL} intensity map (Fig.~\ref{MoS2on5nm}\textbf{(b)}), and the average \gls{CL} spectra extracted from the hyperspectral map using the segmentation mask (Fig.~\ref{MoS2on5nm}\textbf{(c)}).

\begin{figure}[h!]
	\centering
	\includegraphics[width=1\textwidth]{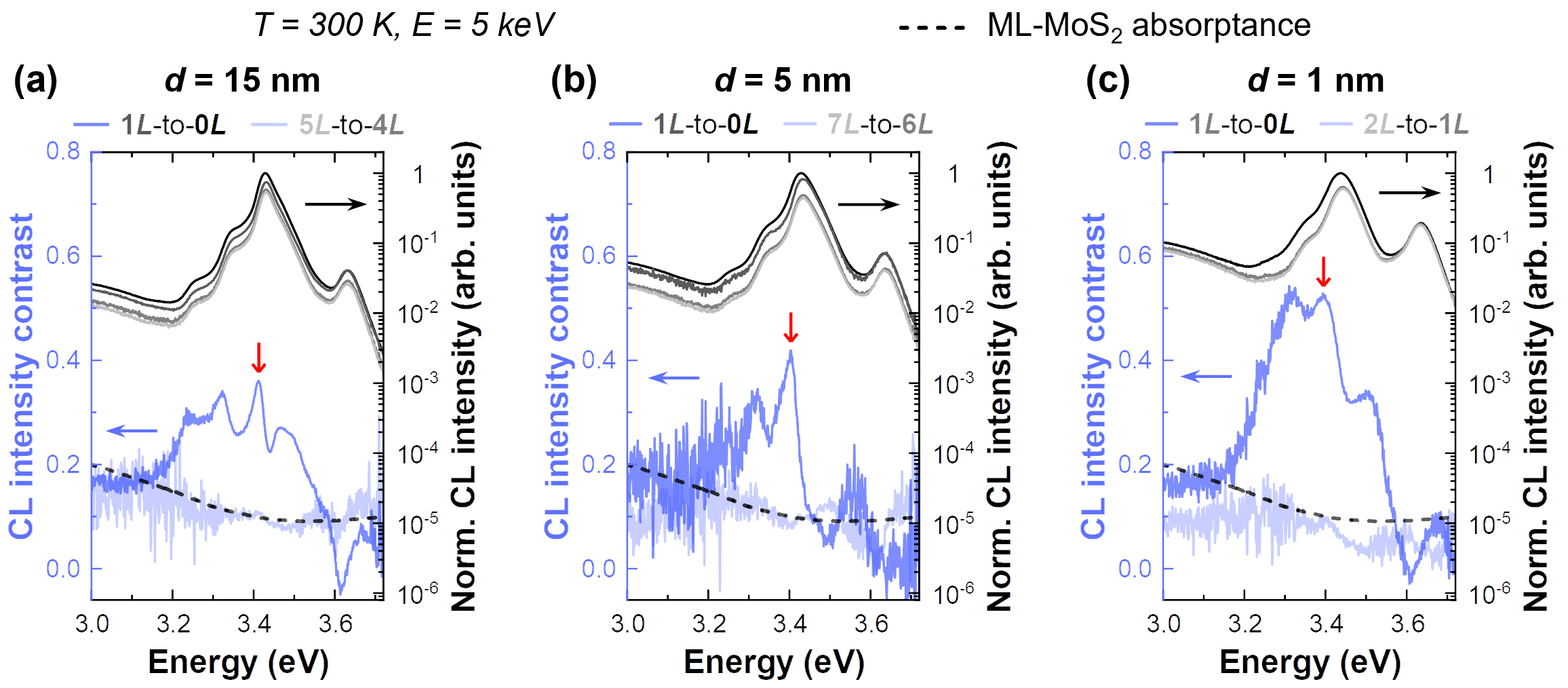}
	\caption{
		The blue curves represent the \gls{CL} intensity contrast between the background region ($0L$) and the region coated with \gls{1L}-MoS\textsubscript{2} ($1L$).
		The light blue curves represent the \gls{CL} intensity contrast between regions coated with $n$ and $(n+1)$ \glspl{1L} of MoS\textsubscript{2}, where $n>0$.
		The black dashed curves represent the reported spectral absorptance of \gls{1L}-MoS\textsubscript{2} \cite{Dumcenco2015}.
		The data is obtained from surface GaN \glspl{QW} with \textbf{(a)} $d=15$~nm, \textbf{(b)} $d=5$~nm, and \textbf{(c)} $d=1$~nm.
		The related \gls{CL} spectra are plotted to illustrate the energy range of different emissions.
		The colors of the spectra correspond to the colors used for their labels, $nL$, representing the number of MoS\textsubscript{2} \glspl{1L} in the corresponding region.
	}
	\label{Contrasts}
\end{figure}

By extracting average spectra from regions with varying MoS\textsubscript{2} thicknesses across all the samples in the series, we calculated the \gls{CL} intensity contrast between the background ($0L$) and the region coated with \gls{1L}-MoS\textsubscript{2} ($1L$), as well as between two other regions with MoS\textsubscript{2} thickness differing by 1~\gls{1L} ($nL$ and $(n+1)L$, with $n>0$).
As shown in Fig.~\ref{Contrasts}, the contrast observed in other regions, induced by the presence of the additional \gls{1L}-MoS\textsubscript{2} ($(n+1)L$-to-$nL$, light blue curve) aligns with \gls{1L}-MoS\textsubscript{2} spectral absorptance \cite{Dumcenco2015} (black dashed curve).
However, the contrast observed on the \gls{QW} surface, resulting from the deposition of a single \gls{1L}-MoS\textsubscript{2} ($1L$-to-$0L$, blue curve), differs significantly from the absorption behavior and displays distinct peaks associated with the \gls{QW} emissions.
It is worth noting that the shape of these contrast peaks varies between different samples, which is likely due to variations in excitonic transitions within the \glspl{QW} of different $d$, influenced by structural or material disorders that arise during the growth process.
Let us focus on the peak around 3.4~eV (indicated by the red arrow), which corresponds to the \gls{ZPL} of the \glspl{QW}.
The peak contrast is $\sim36\%$ for $d=15$~nm (Fig.~\ref{Contrasts}\textbf{(a)}), $\sim42\%$ for $d=5$~nm (Fig.~\ref{Contrasts}\textbf{(b)}), and $\sim52\%$ for $d=1$~nm (Fig.~\ref{Contrasts}\textbf{(c)}).
The amplitude of this peak reflects the strength of the excitonic interactions between the GaN \gls{QW} and \gls{1L}-MoS\textsubscript{2}. 
It is evident that the interaction strength is strongly dependent on the distance between the two materials, \latin{i.e.}, $d$. 
This observation is in line with the mechanism of \gls{FRET} \cite{Forster1960}.

\clearpage

\pdfbookmark[0]{References}{sec:ref}
\bibliography{References}

\providecommand{\latin}[1]{#1}
\makeatletter
\providecommand{\doi}
  {\begingroup\let\do\@makeother\dospecials
  \catcode`\{=1 \catcode`\}=2 \doi@aux}
\providecommand{\doi@aux}[1]{\endgroup\texttt{#1}}
\makeatother
\providecommand*\mcitethebibliography{\thebibliography}
\csname @ifundefined\endcsname{endmcitethebibliography}
  {\let\endmcitethebibliography\endthebibliography}{}
\begin{mcitethebibliography}{56}
\providecommand*\natexlab[1]{#1}
\providecommand*\mciteSetBstSublistMode[1]{}
\providecommand*\mciteSetBstMaxWidthForm[2]{}
\providecommand*\mciteBstWouldAddEndPuncttrue
  {\def\EndOfBibitem{\unskip.}}
\providecommand*\mciteBstWouldAddEndPunctfalse
  {\let\EndOfBibitem\relax}
\providecommand*\mciteSetBstMidEndSepPunct[3]{}
\providecommand*\mciteSetBstSublistLabelBeginEnd[3]{}
\providecommand*\EndOfBibitem{}
\mciteSetBstSublistMode{f}
\mciteSetBstMaxWidthForm{subitem}{(\alph{mcitesubitemcount})}
\mciteSetBstSublistLabelBeginEnd
  {\mcitemaxwidthsubitemform\space}
  {\relax}
  {\relax}

\bibitem[Nakamura \latin{et~al.}(1994)Nakamura, Mukai, and Senoh]{Nakamura1994}
Nakamura,~S.; Mukai,~T.; Senoh,~M. Candela-class high-brightness InGaN/AlGaN
  double-heterostructure blue-light-emitting diodes. \emph{Appl. Phys. Lett.}
  \textbf{1994}, \emph{64}, 1687--1689\relax
\mciteBstWouldAddEndPuncttrue
\mciteSetBstMidEndSepPunct{\mcitedefaultmidpunct}
{\mcitedefaultendpunct}{\mcitedefaultseppunct}\relax
\EndOfBibitem
\bibitem[Steube \latin{et~al.}(1997)Steube, Reimann, Fröhlich, and
  Clarke]{Steube1997}
Steube,~M.; Reimann,~K.; Fröhlich,~D.; Clarke,~S.~J. Free excitons with $n =
  2$ in bulk GaN. \emph{Appl. Phys. Lett.} \textbf{1997}, \emph{71},
  948--949\relax
\mciteBstWouldAddEndPuncttrue
\mciteSetBstMidEndSepPunct{\mcitedefaultmidpunct}
{\mcitedefaultendpunct}{\mcitedefaultseppunct}\relax
\EndOfBibitem
\bibitem[Malpuech \latin{et~al.}(2002)Malpuech, Di~Carlo, Kavokin, Baumberg,
  Zamfirescu, and Lugli]{Malpuech2002}
Malpuech,~G.; Di~Carlo,~A.; Kavokin,~A.; Baumberg,~J.~J.; Zamfirescu,~M.;
  Lugli,~P. Room-temperature polariton lasers based on GaN microcavities.
  \emph{Appl. Phys. Lett.} \textbf{2002}, \emph{81}, 412--414\relax
\mciteBstWouldAddEndPuncttrue
\mciteSetBstMidEndSepPunct{\mcitedefaultmidpunct}
{\mcitedefaultendpunct}{\mcitedefaultseppunct}\relax
\EndOfBibitem
\bibitem[Christopoulos \latin{et~al.}(2007)Christopoulos, von Högersthal,
  Grundy, Lagoudakis, Kavokin, Baumberg, Christmann, Butté, Feltin, Carlin,
  and Grandjean]{Christopoulos2007}
Christopoulos,~S.; von Högersthal,~G. B.~H.; Grundy,~A. J.~D.;
  Lagoudakis,~P.~G.; Kavokin,~A.~V.; Baumberg,~J.~J.; Christmann,~G.;
  Butté,~R.; Feltin,~E.; Carlin,~J.-F.; Grandjean,~N. Room-temperature
  polariton lasing in semiconductor microcavities. \emph{Phys. Rev. Lett.}
  \textbf{2007}, \emph{98}, 126405\relax
\mciteBstWouldAddEndPuncttrue
\mciteSetBstMidEndSepPunct{\mcitedefaultmidpunct}
{\mcitedefaultendpunct}{\mcitedefaultseppunct}\relax
\EndOfBibitem
\bibitem[Novoselov \latin{et~al.}(2004)Novoselov, Geim, Morozov, Jiang, Zhang,
  Dubonos, Grigorieva, and Firsov]{Novoselov2004}
Novoselov,~K.~S.; Geim,~A.~K.; Morozov,~S.~V.; Jiang,~D.; Zhang,~Y.;
  Dubonos,~S.~V.; Grigorieva,~I.~V.; Firsov,~A.~A. Electric field effect in
  atomically thin carbon films. \emph{Science} \textbf{2004}, \emph{306},
  666--669\relax
\mciteBstWouldAddEndPuncttrue
\mciteSetBstMidEndSepPunct{\mcitedefaultmidpunct}
{\mcitedefaultendpunct}{\mcitedefaultseppunct}\relax
\EndOfBibitem
\bibitem[Chaves \latin{et~al.}(2020)Chaves, Azadani, Alsalman, da~Costa,
  Frisenda, Chaves, Song, Kim, He, Zhou, Castellanos-Gomez, Peeters, Liu,
  Hinkle, Oh, Ye, Koester, Lee, Avouris, Wang, and Low]{Chaves2020}
Chaves,~A. \latin{et~al.}  Bandgap engineering of two-dimensional semiconductor
  materials. \emph{npj 2D Mater. Appl.} \textbf{2020}, \emph{4}, 29\relax
\mciteBstWouldAddEndPuncttrue
\mciteSetBstMidEndSepPunct{\mcitedefaultmidpunct}
{\mcitedefaultendpunct}{\mcitedefaultseppunct}\relax
\EndOfBibitem
\bibitem[Britnell \latin{et~al.}(2013)Britnell, Ribeiro, Eckmann, Jalil, Belle,
  Mishchenko, Kim, Gorbachev, Georgiou, Morozov, Grigorenko, Geim, Casiraghi,
  Neto, and Novoselov]{Britnell2013}
Britnell,~L.; Ribeiro,~R.~M.; Eckmann,~A.; Jalil,~R.; Belle,~B.~D.;
  Mishchenko,~A.; Kim,~Y.-J.; Gorbachev,~R.~V.; Georgiou,~T.; Morozov,~S.~V.;
  Grigorenko,~A.~N.; Geim,~A.~K.; Casiraghi,~C.; Neto,~A. H.~C.;
  Novoselov,~K.~S. Strong light-matter interactions in heterostructures of
  atomically thin films. \emph{Science} \textbf{2013}, \emph{340},
  1311--1314\relax
\mciteBstWouldAddEndPuncttrue
\mciteSetBstMidEndSepPunct{\mcitedefaultmidpunct}
{\mcitedefaultendpunct}{\mcitedefaultseppunct}\relax
\EndOfBibitem
\bibitem[Mueller and Malic(2018)Mueller, and Malic]{Mueller2018}
Mueller,~T.; Malic,~E. Exciton physics and device application of
  two-dimensional transition metal dichalcogenide semiconductors. \emph{npj 2D
  Mater. Appl.} \textbf{2018}, \emph{2}, 29\relax
\mciteBstWouldAddEndPuncttrue
\mciteSetBstMidEndSepPunct{\mcitedefaultmidpunct}
{\mcitedefaultendpunct}{\mcitedefaultseppunct}\relax
\EndOfBibitem
\bibitem[Ciarrocchi \latin{et~al.}(2022)Ciarrocchi, Tagarelli, Avsar, and
  Kis]{Ciarrocchi2022}
Ciarrocchi,~A.; Tagarelli,~F.; Avsar,~A.; Kis,~A. Excitonic devices with van
  der Waals heterostructures: valleytronics meets twistronics. \emph{Nat. Rev.
  Mater.} \textbf{2022}, \emph{7}, 449--464\relax
\mciteBstWouldAddEndPuncttrue
\mciteSetBstMidEndSepPunct{\mcitedefaultmidpunct}
{\mcitedefaultendpunct}{\mcitedefaultseppunct}\relax
\EndOfBibitem
\bibitem[Li \latin{et~al.}(2015)Li, Cheng, Zhou, Wang, Yin, Chen, Weiss, Huang,
  and Duan]{Li2015}
Li,~D.; Cheng,~R.; Zhou,~H.; Wang,~C.; Yin,~A.; Chen,~Y.; Weiss,~N.~O.;
  Huang,~Y.; Duan,~X. Electric-field-induced strong enhancement of
  electroluminescence in multilayer molybdenum disulfide. \emph{Nat. Commun.}
  \textbf{2015}, \emph{6}, 7509\relax
\mciteBstWouldAddEndPuncttrue
\mciteSetBstMidEndSepPunct{\mcitedefaultmidpunct}
{\mcitedefaultendpunct}{\mcitedefaultseppunct}\relax
\EndOfBibitem
\bibitem[Zhang \latin{et~al.}(2018)Zhang, Qian, Li, and Chen]{Zhang2018}
Zhang,~Z.; Qian,~Q.; Li,~B.; Chen,~K.~J. Interface engineering of monolayer
  MoS\textsubscript{2}/GaN hybrid heterostructure: modified band alignment for
  photocatalytic water splitting application by nitridation treatment.
  \emph{ACS Appl. Mater. Interfaces} \textbf{2018}, \emph{10},
  17419--17426\relax
\mciteBstWouldAddEndPuncttrue
\mciteSetBstMidEndSepPunct{\mcitedefaultmidpunct}
{\mcitedefaultendpunct}{\mcitedefaultseppunct}\relax
\EndOfBibitem
\bibitem[Jain \latin{et~al.}(2020)Jain, Kumar, Aggarwal, Vashishtha, Goswami,
  Kuriakose, Pandey, Bhaskaran, Walia, and Gupta]{Jain2020}
Jain,~S.~K.; Kumar,~R.~R.; Aggarwal,~N.; Vashishtha,~P.; Goswami,~L.;
  Kuriakose,~S.; Pandey,~A.; Bhaskaran,~M.; Walia,~S.; Gupta,~G. Current
  transport and band alignment study of MoS\textsubscript{2}/GaN and
  MoS\textsubscript{2}/AlGaN heterointerfaces for broadband photodetection
  application. \emph{ACS Appl. Electron. Mater.} \textbf{2020}, \emph{2},
  710--718\relax
\mciteBstWouldAddEndPuncttrue
\mciteSetBstMidEndSepPunct{\mcitedefaultmidpunct}
{\mcitedefaultendpunct}{\mcitedefaultseppunct}\relax
\EndOfBibitem
\bibitem[Chang \latin{et~al.}(1993)Chang, Tan, Zhang, Bimberg, Merz, and
  Hu]{Chang1993}
Chang,~Y.-L.; Tan,~I.-H.; Zhang,~Y.-H.; Bimberg,~D.; Merz,~J.; Hu,~E. Reduced
  quantum efficiency of a near-surface quantum well. \emph{J. Appl. Phys.}
  \textbf{1993}, \emph{74}, 5144--5148\relax
\mciteBstWouldAddEndPuncttrue
\mciteSetBstMidEndSepPunct{\mcitedefaultmidpunct}
{\mcitedefaultendpunct}{\mcitedefaultseppunct}\relax
\EndOfBibitem
\bibitem[Bulashevich and Karpov(2016)Bulashevich, and Karpov]{Karpov2016}
Bulashevich,~K.~A.; Karpov,~S.~Y. Impact of surface recombination on efficiency
  of III-nitride light-emitting diodes. \emph{Phys. Status Solidi RRL}
  \textbf{2016}, \emph{10}, 480--484\relax
\mciteBstWouldAddEndPuncttrue
\mciteSetBstMidEndSepPunct{\mcitedefaultmidpunct}
{\mcitedefaultendpunct}{\mcitedefaultseppunct}\relax
\EndOfBibitem
\bibitem[Grandjean \latin{et~al.}(1999)Grandjean, Damilano, Dalmasso, Leroux,
  Laügt, and Massies]{Grandjean1999}
Grandjean,~N.; Damilano,~B.; Dalmasso,~S.; Leroux,~M.; Laügt,~M.; Massies,~J.
  Built-in electric-field effects in wurtzite AlGaN/GaN quantum wells. \emph{J.
  Appl. Phys.} \textbf{1999}, \emph{86}, 3714--3720\relax
\mciteBstWouldAddEndPuncttrue
\mciteSetBstMidEndSepPunct{\mcitedefaultmidpunct}
{\mcitedefaultendpunct}{\mcitedefaultseppunct}\relax
\EndOfBibitem
\bibitem[Bernardini \latin{et~al.}(1997)Bernardini, Fiorentini, and
  Vanderbilt]{Bernardini1997}
Bernardini,~F.; Fiorentini,~V.; Vanderbilt,~D. Spontaneous polarization and
  piezoelectric constants of III-V nitrides. \emph{Phys. Rev. B} \textbf{1997},
  \emph{56}, R10024--R10027\relax
\mciteBstWouldAddEndPuncttrue
\mciteSetBstMidEndSepPunct{\mcitedefaultmidpunct}
{\mcitedefaultendpunct}{\mcitedefaultseppunct}\relax
\EndOfBibitem
\bibitem[Förster(1960)]{Forster1960}
Förster,~T. Transfer mechanisms of electronic excitation energy. \emph{Radiat.
  Res. Suppl.} \textbf{1960}, \emph{2}, 326--339\relax
\mciteBstWouldAddEndPuncttrue
\mciteSetBstMidEndSepPunct{\mcitedefaultmidpunct}
{\mcitedefaultendpunct}{\mcitedefaultseppunct}\relax
\EndOfBibitem
\bibitem[Achermann \latin{et~al.}(2004)Achermann, Petruska, Kos, Smith,
  Koleske, and Klimov]{Achermann2004}
Achermann,~M.; Petruska,~M.~A.; Kos,~S.; Smith,~D.~L.; Koleske,~D.~D.;
  Klimov,~V.~I. Energy-transfer pumping of semiconductor nanocrystals using an
  epitaxial quantum well. \emph{Nature} \textbf{2004}, \emph{429},
  642--646\relax
\mciteBstWouldAddEndPuncttrue
\mciteSetBstMidEndSepPunct{\mcitedefaultmidpunct}
{\mcitedefaultendpunct}{\mcitedefaultseppunct}\relax
\EndOfBibitem
\bibitem[Gonzalez \latin{et~al.}(2001)Gonzalez, Bunker, and
  Russell]{Gonzalez2001}
Gonzalez,~J.~C.; Bunker,~K.~L.; Russell,~P.~E. Minority-carrier diffusion
  length in a GaN-based light-emitting diode. \emph{Appl. Phys. Lett.}
  \textbf{2001}, \emph{79}, 1567–1569\relax
\mciteBstWouldAddEndPuncttrue
\mciteSetBstMidEndSepPunct{\mcitedefaultmidpunct}
{\mcitedefaultendpunct}{\mcitedefaultseppunct}\relax
\EndOfBibitem
\bibitem[Van~de Walle and Segev(2007)Van~de Walle, and Segev]{VdW2007}
Van~de Walle,~C.~G.; Segev,~D. Microscopic origins of surface states on nitride
  surfaces. \emph{J. Appl. Phys.} \textbf{2007}, \emph{101}, 081704\relax
\mciteBstWouldAddEndPuncttrue
\mciteSetBstMidEndSepPunct{\mcitedefaultmidpunct}
{\mcitedefaultendpunct}{\mcitedefaultseppunct}\relax
\EndOfBibitem
\bibitem[Seong and Amano(2020)Seong, and Amano]{Seong2020}
Seong,~T.-Y.; Amano,~H. Surface passivation of light emitting diodes: From
  nano-size to conventional mesa-etched devices. \emph{Surf. Interfaces}
  \textbf{2020}, \emph{21}, 100765\relax
\mciteBstWouldAddEndPuncttrue
\mciteSetBstMidEndSepPunct{\mcitedefaultmidpunct}
{\mcitedefaultendpunct}{\mcitedefaultseppunct}\relax
\EndOfBibitem
\bibitem[Li \latin{et~al.}(2013)Li, Wu, Huang, Lu, Yang, Lu, Xiong, and
  Zhang]{Li2013}
Li,~H.; Wu,~J.; Huang,~X.; Lu,~G.; Yang,~J.; Lu,~X.; Xiong,~Q.; Zhang,~H. Rapid
  and reliable thickness identification of two-dimensional nanosheets using
  optical microscopy. \emph{ACS Nano} \textbf{2013}, \emph{7},
  10344--10353\relax
\mciteBstWouldAddEndPuncttrue
\mciteSetBstMidEndSepPunct{\mcitedefaultmidpunct}
{\mcitedefaultendpunct}{\mcitedefaultseppunct}\relax
\EndOfBibitem
\bibitem[Dumcenco \latin{et~al.}(2015)Dumcenco, Ovchinnikov, Marinov, Lazić,
  Gibertini, Marzari, Sanchez, Kung, Krasnozhon, Chen, Bertolazzi, Gillet,
  Fontcuberta~i Morral, Radenovic, and Kis]{Dumcenco2015}
Dumcenco,~D.; Ovchinnikov,~D.; Marinov,~K.; Lazić,~P.; Gibertini,~M.;
  Marzari,~N.; Sanchez,~O.~L.; Kung,~Y.-C.; Krasnozhon,~D.; Chen,~M.-W.;
  Bertolazzi,~S.; Gillet,~P.; Fontcuberta~i Morral,~A.; Radenovic,~A.; Kis,~A.
  Large-area epitaxial monolayer MoS\textsubscript{2}. \emph{ACS Nano}
  \textbf{2015}, \emph{9}, 4611–4620\relax
\mciteBstWouldAddEndPuncttrue
\mciteSetBstMidEndSepPunct{\mcitedefaultmidpunct}
{\mcitedefaultendpunct}{\mcitedefaultseppunct}\relax
\EndOfBibitem
\bibitem[Castellanos-Gomez \latin{et~al.}(2016)Castellanos-Gomez, Quereda,
  van~der Meulen, Agraït, and Rubio-Bollinger]{Castellanos-Gomez2016}
Castellanos-Gomez,~A.; Quereda,~J.; van~der Meulen,~H.~P.; Agraït,~N.;
  Rubio-Bollinger,~G. Spatially resolved optical absorption spectroscopy of
  single- and few-layer MoS\textsubscript{2} by hyperspectral imaging.
  \emph{Nanotechnology} \textbf{2016}, \emph{27}, 115705\relax
\mciteBstWouldAddEndPuncttrue
\mciteSetBstMidEndSepPunct{\mcitedefaultmidpunct}
{\mcitedefaultendpunct}{\mcitedefaultseppunct}\relax
\EndOfBibitem
\bibitem[Negri \latin{et~al.}(2020)Negri, Francaviglia, Dumcenco, Bosi, Kaplan,
  Swaminathan, Salviati, Kis, Fabbri, and Fontcuberta~i Morral]{Negri2020}
Negri,~M.; Francaviglia,~L.; Dumcenco,~D.; Bosi,~M.; Kaplan,~D.;
  Swaminathan,~V.; Salviati,~G.; Kis,~A.; Fabbri,~F.; Fontcuberta~i Morral,~A.
  Quantitative nanoscale absorption mapping: a novel technique to probe optical
  absorption of two-dimensional materials. \emph{Nano Lett.} \textbf{2020},
  \emph{20}, 567--576\relax
\mciteBstWouldAddEndPuncttrue
\mciteSetBstMidEndSepPunct{\mcitedefaultmidpunct}
{\mcitedefaultendpunct}{\mcitedefaultseppunct}\relax
\EndOfBibitem
\bibitem[Rossbach \latin{et~al.}(2014)Rossbach, Levrat, Jacopin, Shahmohammadi,
  Carlin, Gani\`ere, Butt\'e, Deveaud, and Grandjean]{Rossbach2014}
Rossbach,~G.; Levrat,~J.; Jacopin,~G.; Shahmohammadi,~M.; Carlin,~J.-F.;
  Gani\`ere,~J.-D.; Butt\'e,~R.; Deveaud,~B.; Grandjean,~N. High-temperature
  Mott transition in wide-band-gap semiconductor quantum wells. \emph{Phys.
  Rev. B} \textbf{2014}, \emph{90}, 201308\relax
\mciteBstWouldAddEndPuncttrue
\mciteSetBstMidEndSepPunct{\mcitedefaultmidpunct}
{\mcitedefaultendpunct}{\mcitedefaultseppunct}\relax
\EndOfBibitem
\bibitem[Shahmohammadi \latin{et~al.}(2014)Shahmohammadi, Jacopin, Rossbach,
  Levrat, Feltin, Carlin, Ganière, Butté, Grandjean, and
  Deveaud]{Shahmohammadi2014}
Shahmohammadi,~M.; Jacopin,~G.; Rossbach,~G.; Levrat,~J.; Feltin,~E.;
  Carlin,~J.-F.; Ganière,~J.-D.; Butté,~R.; Grandjean,~N.; Deveaud,~B.
  Biexcitonic molecules survive excitons at the Mott transition. \emph{Nat.
  Commun.} \textbf{2014}, \emph{5}, 5251\relax
\mciteBstWouldAddEndPuncttrue
\mciteSetBstMidEndSepPunct{\mcitedefaultmidpunct}
{\mcitedefaultendpunct}{\mcitedefaultseppunct}\relax
\EndOfBibitem
\bibitem[Klingshirn(2005)]{Klingshirn2005}
Klingshirn,~C.~F. \emph{Semiconductor optics}; Springer: Berlin; New York,
  2005\relax
\mciteBstWouldAddEndPuncttrue
\mciteSetBstMidEndSepPunct{\mcitedefaultmidpunct}
{\mcitedefaultendpunct}{\mcitedefaultseppunct}\relax
\EndOfBibitem
\bibitem[Hofstetter \latin{et~al.}(2020)Hofstetter, Beck, Epler, Kirste, and
  Bour]{Hofstetter2020}
Hofstetter,~D.; Beck,~H.; Epler,~J.~E.; Kirste,~L.; Bour,~D.~P. Evidence of
  strong electron-phonon interaction in a GaN-based quantum cascade emitter.
  \emph{Superlattices Microstruct.} \textbf{2020}, \emph{145}, 106631\relax
\mciteBstWouldAddEndPuncttrue
\mciteSetBstMidEndSepPunct{\mcitedefaultmidpunct}
{\mcitedefaultendpunct}{\mcitedefaultseppunct}\relax
\EndOfBibitem
\bibitem[Itskos \latin{et~al.}(2007)Itskos, Heliotis, Lagoudakis, Lupton,
  Barradas, Alves, Pereira, Watson, Dawson, Feldmann, Murray, and
  Bradley]{Itskos2007}
Itskos,~G.; Heliotis,~G.; Lagoudakis,~P.~G.; Lupton,~J.; Barradas,~N.~P.;
  Alves,~E.; Pereira,~S.; Watson,~I.~M.; Dawson,~M.~D.; Feldmann,~J.;
  Murray,~R.; Bradley,~D. D.~C. Efficient dipole-dipole coupling of
  Mott-Wannier and Frenkel excitons in (Ga,In)N quantum well/polyfluorene
  semiconductor heterostructures. \emph{Phys. Rev. B} \textbf{2007}, \emph{76},
  035344\relax
\mciteBstWouldAddEndPuncttrue
\mciteSetBstMidEndSepPunct{\mcitedefaultmidpunct}
{\mcitedefaultendpunct}{\mcitedefaultseppunct}\relax
\EndOfBibitem
\bibitem[Prins \latin{et~al.}(2014)Prins, Goodman, and Tisdale]{Prins2014}
Prins,~F.; Goodman,~A.~J.; Tisdale,~W.~A. Reduced dielectric screening and
  enhanced energy transfer in single- and few-layer MoS\textsubscript{2}.
  \emph{Nano Lett.} \textbf{2014}, \emph{14}, 6087--6091\relax
\mciteBstWouldAddEndPuncttrue
\mciteSetBstMidEndSepPunct{\mcitedefaultmidpunct}
{\mcitedefaultendpunct}{\mcitedefaultseppunct}\relax
\EndOfBibitem
\bibitem[Taghipour \latin{et~al.}(2018)Taghipour, Hernandez~Martinez, Ozden,
  Olutas, Dede, Gungor, Erdem, Perkgoz, and Demir]{Taghipour2018}
Taghipour,~N.; Hernandez~Martinez,~P.~L.; Ozden,~A.; Olutas,~M.; Dede,~D.;
  Gungor,~K.; Erdem,~O.; Perkgoz,~N.~K.; Demir,~H.~V. Near-unity efficiency
  energy transfer from colloidal semiconductor quantum wells of CdSe/CdS
  nanoplatelets to a monolayer of MoS\textsubscript{2}. \emph{ACS Nano}
  \textbf{2018}, \emph{12}, 8547--8554\relax
\mciteBstWouldAddEndPuncttrue
\mciteSetBstMidEndSepPunct{\mcitedefaultmidpunct}
{\mcitedefaultendpunct}{\mcitedefaultseppunct}\relax
\EndOfBibitem
\bibitem[Feltin \latin{et~al.}(2007)Feltin, Simeonov, Carlin, Butt\'e, and
  Grandjean]{Feltin2007}
Feltin,~E.; Simeonov,~D.; Carlin,~J.-F.; Butt\'e,~R.; Grandjean,~N. Narrow UV
  emission from homogeneous GaN/AlGaN quantum wells. \emph{Appl. Phys. Lett.}
  \textbf{2007}, \emph{90}, 021905\relax
\mciteBstWouldAddEndPuncttrue
\mciteSetBstMidEndSepPunct{\mcitedefaultmidpunct}
{\mcitedefaultendpunct}{\mcitedefaultseppunct}\relax
\EndOfBibitem
\bibitem[Wang \latin{et~al.}(2013)Wang, Meric, Huang, Gao, Gao, Tran,
  Taniguchi, Watanabe, Campos, Muller, Guo, Kim, Hone, Shepard, and
  Dean]{Wang2013}
Wang,~L.; Meric,~I.; Huang,~P.~Y.; Gao,~Q.; Gao,~Y.; Tran,~H.; Taniguchi,~T.;
  Watanabe,~K.; Campos,~L.~M.; Muller,~D.~A.; Guo,~J.; Kim,~P.; Hone,~J.;
  Shepard,~K.~L.; Dean,~C.~R. One-dimensional electrical contact to a
  two-dimensional material. \emph{Science} \textbf{2013}, \emph{342},
  614--617\relax
\mciteBstWouldAddEndPuncttrue
\mciteSetBstMidEndSepPunct{\mcitedefaultmidpunct}
{\mcitedefaultendpunct}{\mcitedefaultseppunct}\relax
\EndOfBibitem
\bibitem[Castellanos-Gomez \latin{et~al.}(2022)Castellanos-Gomez, Duan, Fei,
  Gutierrez, Huang, Huang, Quereda, Qian, Sutter, and
  Sutter]{Castellanos-Gomez2022}
Castellanos-Gomez,~A.; Duan,~X.; Fei,~Z.; Gutierrez,~H.~R.; Huang,~Y.;
  Huang,~X.; Quereda,~J.; Qian,~Q.; Sutter,~E.; Sutter,~P. Van der Waals
  heterostructures. \emph{Nat. Rev. Methods Primers} \textbf{2022}, \emph{2},
  58\relax
\mciteBstWouldAddEndPuncttrue
\mciteSetBstMidEndSepPunct{\mcitedefaultmidpunct}
{\mcitedefaultendpunct}{\mcitedefaultseppunct}\relax
\EndOfBibitem
\bibitem[Drouin \latin{et~al.}(2007)Drouin, Couture, Joly, Tastet, Aimez, and
  Gauvin]{Drouin2007}
Drouin,~D.; Couture,~A.~R.; Joly,~D.; Tastet,~X.; Aimez,~V.; Gauvin,~R. CASINO
  V2.42 — A fast and easy-to-use modeling tool for scanning electron
  microscopy and microanalysis users. \emph{Scanning} \textbf{2007}, \emph{29},
  92--101\relax
\mciteBstWouldAddEndPuncttrue
\mciteSetBstMidEndSepPunct{\mcitedefaultmidpunct}
{\mcitedefaultendpunct}{\mcitedefaultseppunct}\relax
\EndOfBibitem
\bibitem[Evoy \latin{et~al.}(1999)Evoy, Harnett, Keller, Mishra, DenBaars, and
  Craighead]{Evoy2000}
Evoy,~S.; Harnett,~C.~K.; Keller,~S.; Mishra,~U.~K.; DenBaars,~S.~P.;
  Craighead,~H.~G. Scanning tunneling microscope-induced luminescence studies
  of defects in GaN layers and heterostructures. \emph{MRS Online Proceedings
  Library} \textbf{1999}, \emph{588}, 19\relax
\mciteBstWouldAddEndPuncttrue
\mciteSetBstMidEndSepPunct{\mcitedefaultmidpunct}
{\mcitedefaultendpunct}{\mcitedefaultseppunct}\relax
\EndOfBibitem
\bibitem[Jahn \latin{et~al.}(2022)Jahn, Kaganer, Sabelfeld, Kireeva,
  L\"ahnemann, Pf\"uller, Flissikowski, Ch\`eze, Biermann, Calarco, and
  Brandt]{Jahn2022}
Jahn,~U.; Kaganer,~V.~M.; Sabelfeld,~K.~K.; Kireeva,~A.~E.; L\"ahnemann,~J.;
  Pf\"uller,~C.; Flissikowski,~T.; Ch\`eze,~C.; Biermann,~K.; Calarco,~R.;
  Brandt,~O. Carrier diffusion in $\mathrm{Ga}\mathrm{N}$: A
  cathodoluminescence study. I. Temperature-dependent generation volume.
  \emph{Phys. Rev. Appl.} \textbf{2022}, \emph{17}, 024017\relax
\mciteBstWouldAddEndPuncttrue
\mciteSetBstMidEndSepPunct{\mcitedefaultmidpunct}
{\mcitedefaultendpunct}{\mcitedefaultseppunct}\relax
\EndOfBibitem
\bibitem[Guthrey and Moseley(2020)Guthrey, and Moseley]{Guthrey2020}
Guthrey,~H.; Moseley,~J. A review and perspective on cathodoluminescence
  analysis of halide perovskites. \emph{Adv. Energy Mater.} \textbf{2020},
  \emph{10}, 1903840\relax
\mciteBstWouldAddEndPuncttrue
\mciteSetBstMidEndSepPunct{\mcitedefaultmidpunct}
{\mcitedefaultendpunct}{\mcitedefaultseppunct}\relax
\EndOfBibitem
\bibitem[Brunner \latin{et~al.}(1997)Brunner, Angerer, Bustarret, Freudenberg,
  Höpler, Dimitrov, Ambacher, and Stutzmann]{Brunner1997}
Brunner,~D.; Angerer,~H.; Bustarret,~E.; Freudenberg,~F.; Höpler,~R.;
  Dimitrov,~R.; Ambacher,~O.; Stutzmann,~M. Optical constants of epitaxial
  AlGaN films and their temperature dependence. \emph{J. Appl. Phys.}
  \textbf{1997}, \emph{82}, 5090--5096\relax
\mciteBstWouldAddEndPuncttrue
\mciteSetBstMidEndSepPunct{\mcitedefaultmidpunct}
{\mcitedefaultendpunct}{\mcitedefaultseppunct}\relax
\EndOfBibitem
\bibitem[Podlipskas \latin{et~al.}(2019)Podlipskas, Jurkevi\v{c}ius, Kadys,
  Miasojedovas, Malinauskas, and Aleksiej\={u}nas]{Podlipskas2019}
Podlipskas,~Z.; Jurkevi\v{c}ius,~J.; Kadys,~A.; Miasojedovas,~S.;
  Malinauskas,~T.; Aleksiej\={u}nas,~R. The detrimental effect of AlGaN barrier
  quality on carrier dynamics in AlGaN/GaN interface. \emph{Sci. Rep.}
  \textbf{2019}, \emph{9}, 17346\relax
\mciteBstWouldAddEndPuncttrue
\mciteSetBstMidEndSepPunct{\mcitedefaultmidpunct}
{\mcitedefaultendpunct}{\mcitedefaultseppunct}\relax
\EndOfBibitem
\bibitem[Liu(2019)]{Liu2019}
Liu,~W. \emph{Ultrafast carrier dynamics in III-nitride nanostructures and LED
  quantum efficiency}; Ph.D. Dissertation, \'Ecole Polytechnique F\'ed\'erale
  de Lausanne, Lausanne, Switzerland, 2019\relax
\mciteBstWouldAddEndPuncttrue
\mciteSetBstMidEndSepPunct{\mcitedefaultmidpunct}
{\mcitedefaultendpunct}{\mcitedefaultseppunct}\relax
\EndOfBibitem
\bibitem[Reshchikov \latin{et~al.}(2000)Reshchikov, Shahedipour, Korotkov,
  Wessels, and Ulmer]{Reshchikov2000}
Reshchikov,~M.~A.; Shahedipour,~F.; Korotkov,~R.~Y.; Wessels,~B.~W.;
  Ulmer,~M.~P. Photoluminescence band near 2.9~eV in undoped GaN epitaxial
  layers. \emph{J. Appl. Phys.} \textbf{2000}, \emph{87}, 3351--3354\relax
\mciteBstWouldAddEndPuncttrue
\mciteSetBstMidEndSepPunct{\mcitedefaultmidpunct}
{\mcitedefaultendpunct}{\mcitedefaultseppunct}\relax
\EndOfBibitem
\bibitem[Birner \latin{et~al.}(2007)Birner, Zibold, Andlauer, Kubis, Sabathil,
  Trellakis, and Vogl]{nextnano}
Birner,~S.; Zibold,~T.; Andlauer,~T.; Kubis,~T.; Sabathil,~M.; Trellakis,~A.;
  Vogl,~P. nextnano: General purpose 3-D simulations. \emph{IEEE Trans.
  Electron Devices} \textbf{2007}, \emph{54}, 2137--2142\relax
\mciteBstWouldAddEndPuncttrue
\mciteSetBstMidEndSepPunct{\mcitedefaultmidpunct}
{\mcitedefaultendpunct}{\mcitedefaultseppunct}\relax
\EndOfBibitem
\bibitem[Rosencher and Vinter(2002)Rosencher, and Vinter]{Rosencher2002}
Rosencher,~E.; Vinter,~B. \emph{Optoelectronics}; Cambridge University Press:
  Cambridge, UK, 2002\relax
\mciteBstWouldAddEndPuncttrue
\mciteSetBstMidEndSepPunct{\mcitedefaultmidpunct}
{\mcitedefaultendpunct}{\mcitedefaultseppunct}\relax
\EndOfBibitem
\bibitem[Vurgaftman and Meyer(2003)Vurgaftman, and Meyer]{Vurgaftman2003}
Vurgaftman,~I.; Meyer,~J.~R. Band parameters for nitrogen-containing
  semiconductors. \emph{J. Appl. Phys.} \textbf{2003}, \emph{94},
  3675--3696\relax
\mciteBstWouldAddEndPuncttrue
\mciteSetBstMidEndSepPunct{\mcitedefaultmidpunct}
{\mcitedefaultendpunct}{\mcitedefaultseppunct}\relax
\EndOfBibitem
\bibitem[Nath \latin{et~al.}(2013)Nath, Yang, Lee, Park, Wu, and
  Rajan]{Nath2013}
Nath,~D.~N.; Yang,~Z.~C.; Lee,~C.-Y.; Park,~P.~S.; Wu,~Y.-R.; Rajan,~S.
  Unipolar vertical transport in GaN/AlGaN/GaN heterostructures. \emph{Appl.
  Phys. Lett.} \textbf{2013}, \emph{103}, 022102\relax
\mciteBstWouldAddEndPuncttrue
\mciteSetBstMidEndSepPunct{\mcitedefaultmidpunct}
{\mcitedefaultendpunct}{\mcitedefaultseppunct}\relax
\EndOfBibitem
\bibitem[Li \latin{et~al.}(2017)Li, Piccardo, Lu, Mayboroda, Martinelli,
  Peretti, Speck, Weisbuch, Filoche, and Wu]{Li2017}
Li,~C.-K.; Piccardo,~M.; Lu,~L.-S.; Mayboroda,~S.; Martinelli,~L.; Peretti,~J.;
  Speck,~J.~S.; Weisbuch,~C.; Filoche,~M.; Wu,~Y.-R. Localization landscape
  theory of disorder in semiconductors. III. Application to carrier transport
  and recombination in light emitting diodes. \emph{Phys. Rev. B}
  \textbf{2017}, \emph{95}, 144206\relax
\mciteBstWouldAddEndPuncttrue
\mciteSetBstMidEndSepPunct{\mcitedefaultmidpunct}
{\mcitedefaultendpunct}{\mcitedefaultseppunct}\relax
\EndOfBibitem
\bibitem[Trivi\~{n}o \latin{et~al.}(2015)Trivi\~{n}o, Butt\'{e}, Carlin, and
  Grandjean]{Trivino2015}
Trivi\~{n}o,~N.~V.; Butt\'{e},~R.; Carlin,~J.-F.; Grandjean,~N. Continuous wave
  blue lasing in III-nitride nanobeam cavity on silicon. \emph{Nano Lett.}
  \textbf{2015}, \emph{15}, 1259--1263\relax
\mciteBstWouldAddEndPuncttrue
\mciteSetBstMidEndSepPunct{\mcitedefaultmidpunct}
{\mcitedefaultendpunct}{\mcitedefaultseppunct}\relax
\EndOfBibitem
\bibitem[Benameur \latin{et~al.}(2011)Benameur, Radisavljevic, Héron, Sahoo,
  Berger, and Kis]{Benameur2011}
Benameur,~M.~M.; Radisavljevic,~B.; Héron,~J.~S.; Sahoo,~S.; Berger,~H.;
  Kis,~A. Visibility of dichalcogenide nanolayers. \emph{Nanotechnology}
  \textbf{2011}, \emph{22}, 125706\relax
\mciteBstWouldAddEndPuncttrue
\mciteSetBstMidEndSepPunct{\mcitedefaultmidpunct}
{\mcitedefaultendpunct}{\mcitedefaultseppunct}\relax
\EndOfBibitem
\bibitem[Lee \latin{et~al.}(2010)Lee, Yan, Brus, Heinz, Hone, and Ryu]{Lee2010}
Lee,~C.; Yan,~H.; Brus,~L.~E.; Heinz,~T.~F.; Hone,~J.; Ryu,~S. Anomalous
  lattice vibrations of single- and few-layer MoS\textsubscript{2}. \emph{ACS
  Nano} \textbf{2010}, \emph{4}, 2695--2700\relax
\mciteBstWouldAddEndPuncttrue
\mciteSetBstMidEndSepPunct{\mcitedefaultmidpunct}
{\mcitedefaultendpunct}{\mcitedefaultseppunct}\relax
\EndOfBibitem
\bibitem[Nemes-Incze \latin{et~al.}(2008)Nemes-Incze, Osv\'ath, Kamar\'as, and
  Bir\'o]{Nemes-Incze2008}
Nemes-Incze,~P.; Osv\'ath,~Z.; Kamar\'as,~K.; Bir\'o,~L.~P. Anomalies in
  thickness measurements of graphene and few layer graphite crystals by tapping
  mode atomic force microscopy. \emph{Carbon} \textbf{2008}, \emph{46},
  1435--1442\relax
\mciteBstWouldAddEndPuncttrue
\mciteSetBstMidEndSepPunct{\mcitedefaultmidpunct}
{\mcitedefaultendpunct}{\mcitedefaultseppunct}\relax
\EndOfBibitem
\bibitem[Li \latin{et~al.}(2012)Li, Zhang, Yap, Tay, Edwin, Olivier, and
  Baillargeat]{Li2012}
Li,~H.; Zhang,~Q.; Yap,~C. C.~R.; Tay,~B.~K.; Edwin,~T. H.~T.; Olivier,~A.;
  Baillargeat,~D. From bulk to monolayer MoS\textsubscript{2}: Evolution of
  Raman scattering. \emph{Adv. Funct. Mater.} \textbf{2012}, \emph{22},
  1385--1390\relax
\mciteBstWouldAddEndPuncttrue
\mciteSetBstMidEndSepPunct{\mcitedefaultmidpunct}
{\mcitedefaultendpunct}{\mcitedefaultseppunct}\relax
\EndOfBibitem
\bibitem[Buscema \latin{et~al.}(2014)Buscema, Steele, van~der Zant, and
  Castellanos-Gomez]{Buscema2014}
Buscema,~M.; Steele,~G.~A.; van~der Zant,~H. S.~J.; Castellanos-Gomez,~A. The
  effect of the substrate on the Raman and photoluminescence emission of
  single-layer MoS\textsubscript{2}. \emph{Nano Res.} \textbf{2014}, \emph{7},
  561--571\relax
\mciteBstWouldAddEndPuncttrue
\mciteSetBstMidEndSepPunct{\mcitedefaultmidpunct}
{\mcitedefaultendpunct}{\mcitedefaultseppunct}\relax
\EndOfBibitem
\bibitem[de~la Peña \latin{et~al.}(2022)de~la Peña, Prestat, Fauske, Burdet,
  Lähnemann, Jokubauskas, Furnival, Nord, Ostasevicius, MacArthur, Johnstone,
  Sarahan, Taillon, Aarholt, pquinn dls, Migunov, Eljarrat, Caron, Francis,
  Nemoto, Poon, Mazzucco, actions user, Tappy, Cautaerts, Somnath, Slater,
  Walls, Winkler, and Ånes]{francisco_de_la_pena_2022_7263263}
de~la Peña,~F. \latin{et~al.}  hyperspy/hyperspy: Release v1.7.3. 2022;
  \url{https://doi.org/10.5281/zenodo.7263263}\relax
\mciteBstWouldAddEndPuncttrue
\mciteSetBstMidEndSepPunct{\mcitedefaultmidpunct}
{\mcitedefaultendpunct}{\mcitedefaultseppunct}\relax
\EndOfBibitem
\end{mcitethebibliography}

\end{document}